\documentclass[structabstract]{aa}

\usepackage{epsfig}
\usepackage{graphicx}
\usepackage{subfigure}
\usepackage{latexsym}
\usepackage{amssymb}
\usepackage{amsmath}
\usepackage{longtable,multirow}
\usepackage{color}
\usepackage{txfonts}
\usepackage{appendix}
\usepackage{ulem}
\usepackage{footnote}
\usepackage{natbib,twoopt}
\usepackage{booktabs}

\bibstyle{aa}

\begin{document}
\sloppy

\title{{\it Herschel}\thanks{{\it Herschel} is an ESA space observatory with science instruments provided by European-led Principal Investigator consortia and with important participation from NASA.} observations of the Sgr B2 cores:\\ 
Hydrides, warm CO, and cold dust }

\author{M. Etxaluze\inst{1}
  \and J. R. Goicoechea\inst{1}
  \and J. Cernicharo\inst{1}
  \and E. T. Polehampton\inst{2,3}
  \and A. Noriega-Crespo\inst{4} \and\\
  S. Molinari\inst{5}
 \and B. M. Swinyard\inst{2,6}
\and R. Wu\inst{7}
  \and J. Bally\inst{8}}

\institute{Departamento de Astrof\'isica. Centro de Astrobiolog\'ia. CSIC-INTA. Torrej\'on de Ardoz, 28850 Madrid, Spain.\\
\email{etxaluzeam@cab.inta-csic.es}
  \and RAL Space, Rutherford Appleton Laboratory, Oxfordshire, OX11 0QX, UK
  \and Institute for Space Imaging Science, Department of Physics \& Astronomy, University of Lethbridge, Lethbridge, AB T1K3M4, Canada
  \and Spitzer Science Center, 91125 Pasadena, USA
  \and INAF-IFSI, I-00133 Roma, Italy
  \and Dept. of Physics \& Astronomy, University College London, Gower Street, London WC1E 6BT, UK 
 \and Commissariat \`{A} l'\`{E}nergie Atomique, Service d'Astrophysique, Saclay, 91191 Gif-sur-Yvette, France
  \and CASA, University of Colorado, Boulder, USA 80309}
\authorrunning{Etxaluze et al.}

\abstract
{Sagittarius~B2 is one of the most massive and luminous star-forming regions in the Galaxy and shows a very rich chemistry and physical conditions similar to those in much more distant extragalactic starbursts.} 
{We present large-scale far-infrared/submillimeter photometric images and broadband spectroscopic maps taken with the PACS and SPIRE instruments onboard \textit{Herschel}.} 
{High angular resolution dust images (complemented with \textit{Spitzer} MIPS\,24~$\mu$m images) as well as atomic and molecular spectral maps were made and analyzed in order to constrain the dust properties, the gas physical conditions, and the chemical content of this unique region.}
{The spectra towards the Sagittarius~B2 star-forming cores, B2(M) and B2(N), are characterized by strong CO line emission (from $J$=4 to 16), emission lines from high-density tracers (HCN, HCO$^+$, and H$_2$S), [N\,{\sc ii}]\,205$\mu$m  emission from ionized gas, and a large number of absorption lines from light hydride molecules (OH$^+$, H$_2$O$^+$, H$_2$O, CH$^+$, CH, SH$^+$, HF, NH, NH$_2$, and NH$_3$). The rotational population diagrams of CO suggest the presence of two different gas temperature components: an extended warm component with $T_{\rm rot}\sim$50-100 K, which is associated with the extended envelope, and a hotter component at $T_{\rm rot}\sim$200 K  and $T_{\rm rot}\sim$300 K, which is only seen towards the B2(M) and B2(N) cores, respectively. As observed in other Galactic Center clouds, such gas temperatures are significantly higher than the dust temperatures inferred from photometric images ($T_{\rm d}\simeq 20-30$ K). We determined far-IR luminosities ($L_{\rm FIR}{\rm (M)}\sim 5\times10^6$ L$_{\odot}$ and $L_{\rm FIR}{\rm (N)}\sim 1.1\times10^6$ L$_{\odot}$) and total dust masses ($M_{\rm d}{\rm (M)}\sim 2300$ M$_{\odot}$ and $M_{\rm d}{\rm (N)}\sim 2500$ M$_{\odot}$) in the cores. Non-local thermodynamic equilibrium (non-LTE) models of the CO excitation were used to constrain the averaged gas density ($n({\rm H_2})\sim 10^6$ cm$^{-3}$) in the cores (i.e., similar or lower than the critical densities for collisional thermalization of mid- and high-$J$ CO levels). A uniform luminosity ratio, $L{\rm (CO)}/L_{\rm FIR}\sim (1-3)\times 10^{-4}$, is measured along the  extended envelope, suggesting that the same mechanism dominates the heating of the molecular gas at large scales.}
{Sgr~B2 shows extended emission from warm CO gas and cold dust, whereas only the cores show a hotter CO component. The detection of high-density molecular tracers and of strong [N\,{\sc ii}]\,205$\mu$m line emission towards the cores suggests that their morphology must be clumpy to allow UV radiation to escape from the inner H\,{\sc ii} regions. Together with shocks, the strong UV radiation field is likely responsible for the heating of the hot CO component. At larger scales, photodissociation regions (PDR) models can explain both the observed CO line ratios and the uniform $L{\rm (CO)}/L_{\rm FIR}$ luminosity ratios.}


\date{Received; accepted}

\maketitle

\section{Introduction}\label{intro}

	\indent\par{
	Sagittarius B2 (Sgr B2) is a giant molecular cloud at a distance of $\sim$7.1$\pm$1.5 kpc \citep{Kerr86,Ghez08,Gillessen09,Reid90} (from now on, in this paper, the distance to Sgr B2 is assumed to be 8.5 kpc), it is located in the Central Molecular Zone at $\sim$120 pc from the Galactic Center \citep{Lis90} and placed at the semi-major axis of the $\sim$100 pc ring of gas and dust that rotates around the Galactic Center \citep{Molinari11}. Here the gas achieves velocities in the range 50-90 km$\ $s$^{-1}$ \citep{Tsuboi99,Martin90} and molecular hydrogen column densities are as high as $N(\rm {H_2})\sim 10^{25}$ cm$^{-2}$ \citep{Molinari11,Qin11} towards the main star-forming cores.
}
	\par{
	Sgr B2 is one of the most massive molecular clouds in the Galaxy, with a total mass of $M\sim 6\times10^6 M_{\odot}$ \citep{Goldsmith90}, and one of the most luminous star-forming regions, with a total luminosity of $L\sim 10^7 L_{\odot}$ \citep{Goldsmith90,Goldsmith92}. It is also one of the most active star-forming regions in the Galaxy, containing a number of high-mass O-type stars, young stellar objects, and compact H {\sc ii} regions. Star-formation rate is higher than the average in the Milky Way's molecular clouds. Sgr B2 has a dense central component $\sim5-10$ pc in diameter ($\sim2-14$ arcmin), with molecular gas densities $n\left (\rm H_2 \right )\sim 0.3-3\times 10^5$ cm$^{-3}$ \citep{Minh98}. It also contains three main compact cores, Sgr B2(N), Sgr B2(M), and Sgr B2(S) distributed from north to south. The three cores are associated with massive star-formation and present different evolutionary stages. The star-formation activity in the region is suggested to be triggered by the shocks produced by large-scale collisions between molecular clouds \citep{Sato00}, although the main scenario is not fully constrained. The three main cores are surrounded by an extended envelope ($\sim 40$ pc) of lower density, $n\left (\rm H_2 \right )\le 10^4$ cm$^{-3}$, and warm gas, $T_{\rm K}\sim 100$ K \citep{Huttemeister95}. The dominant gas heating mechanisms are not clear yet.
}	
	\par{Sgr B2 presents an exceptional and extremely rich chemistry \citep{Polehampton07}. Many molecules detected towards Sgr B2 have not been detected anywhere else, and many species that are observed in other molecular clouds were first detected in the direction of Sgr B2. It is also one of the best templates to better understand the emission from much more distant unresolved extragalactic starbursts. In fact, the spectrum of Sgr B2 at far-infrared wavelengths presents spectral features similar to those observed in the closest ultraluminous infrared galaxy Arp 220 \citep{Gonzalez04,Goicoechea04} and in the M82 starburst galaxy \citep{Kamenetzky12}. 
\begin{figure}[t]
\resizebox{\hsize}{!}{\includegraphics{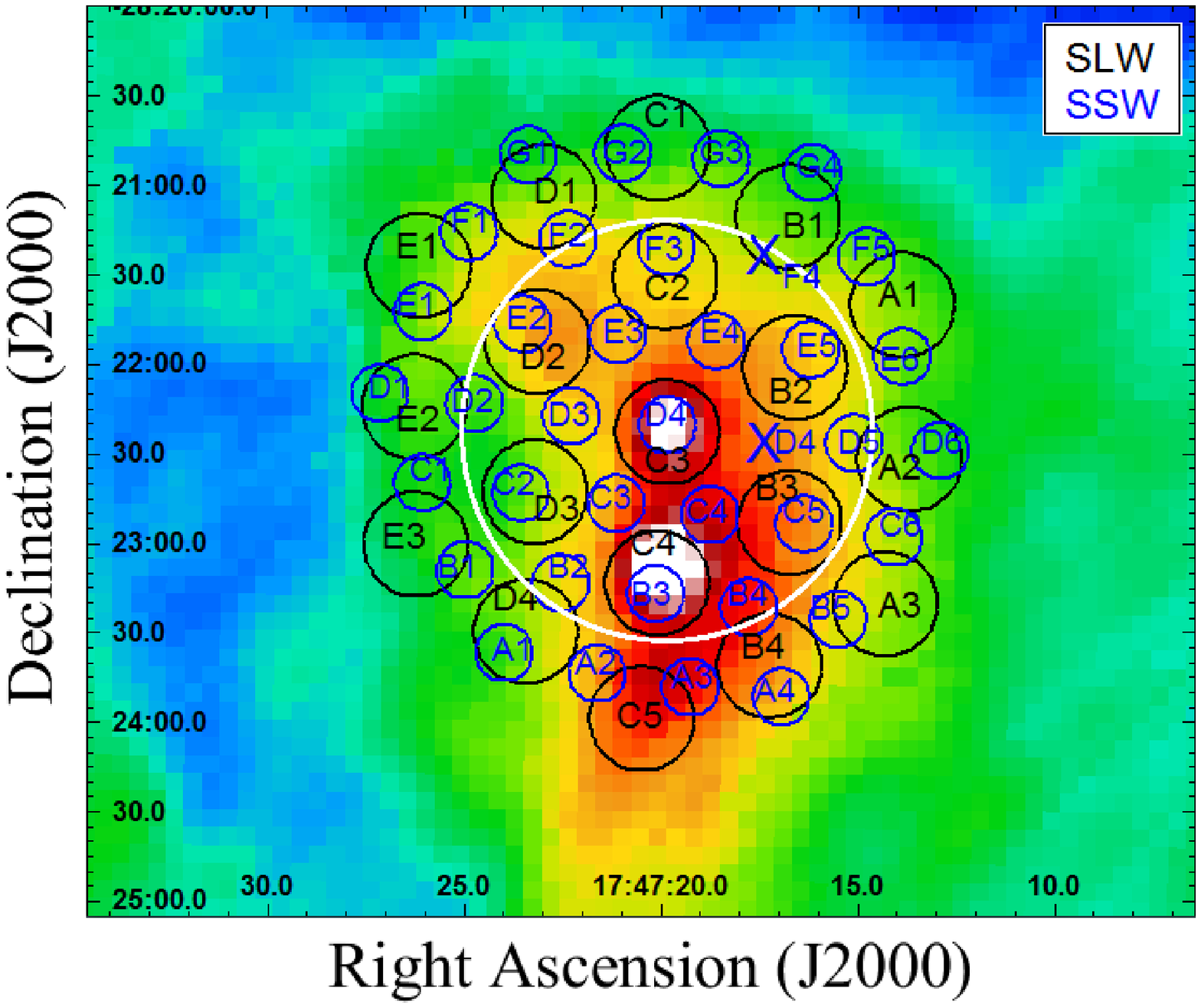}}
\caption{Footprint of the SPIRE FTS detectors (corresponding to the observation obsid:1342214843) over the SPIRE 250 $\mu$m image. Black circles show the layout of the SLW detectors and blue circles show the position of the SSW detectors. Sgr B2(N) is observed with the central detectors SLWC3-SSWD4, while Sgr B2(M) falls onto the detectors SLWC5 and SSWB3. The blue crosses are the two dead detectors in the SSW array (SSWD5 and SSWF4). The white circle represents the FoV with a diameter $\sim$2 arcmin ($\sim$5 pc).} 
\label{fig1}
\end{figure}
}

	\par{The far-infrared continuum observed with the Long Wavelength Spectrometer (LWS) instrument onboard the Infrared Space Observatory (ISO) shows a very strong dust emission through the entire molecular cloud, which is best fit with two dust components: 
a cold component with a temperature of $T_{\rm d}\sim13-22$ K and a warmer component with a temperature $T_{\rm d}\sim 24-38$ K that contributes less than 10 $\%$ to the total hydrogen column density \citep{Goicoechea04}. \citet{Huttemeister95} and \citet{Ceccarelli02} used NH$_3$ lines to show that the gas temperature in the direction of Sgr B2 is high with kinetic temperatures $T_{\rm k}\gtrsim 200$ K.
}
	\par{These gas temperatures are much higher than the values determined for the dust temperature. \citet{Huttemeister95} and \citet{Ceccarelli02} suggested shocks as a possible heating mechanism of the gas. Observations of high-velocity components of HCO$^+$, HNCO and SiO \citep{Minh98,Kuan96,Martin97} towards Sgr B2 are also indicative of strong shocks possibly produced by cloud-cloud collisions. The location of Sgr B2 on the 100 pc ring defined by \citet{Molinari11} coincides with the position where the $x_1$ and the $x_2$ orbits are tangential, implying favorable conditions for cloud-cloud collision. Such a collision may produce strong shocks that trigger the star-formation and alter the chemical composition of the cloud. In addition, \citet{Vastel02} and \citet{Goicoechea04} observed widespread emission of [N\,{\sc II}], [N\,{\sc III}], [O\,{\sc III}], [O\,{\sc I}], and [C\,{\sc II}] fine structure lines over a region of $\sim$35 pc. They suggested a scenario where both shocks and radiative heating mechanisms based on a widespread UV radiation field could explain the heating of the gas to a temperature higher than that of the dust at large scales. Sgr B2 has also been considered an X-ray reflection nebula that shows a diffuse X-ray emission in the K$\alpha$ line of Fe$^0$ at 6.4 keV \citep{Murakami01} with an X-ray luminosity two orders of magnitude larger than the integrated luminosity of all the resolved point sources. \citet{Murakami00} suggested that the region is being irradiated with the X-ray emission that originated in Sgr A$^*$ during a period of high activity in the recent past.
}
\begin{figure}[t]
\resizebox{\hsize}{!}{\includegraphics{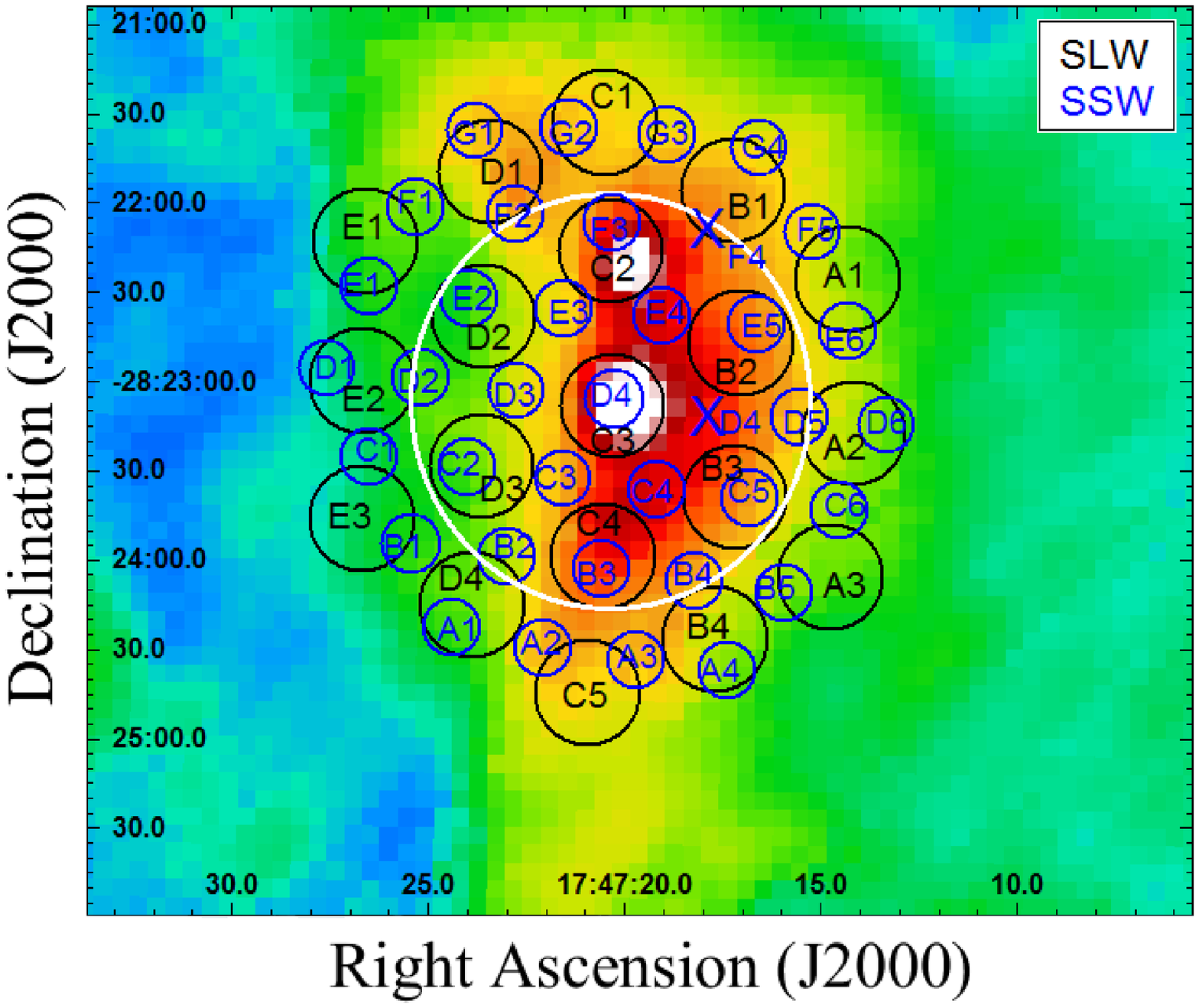}}
\caption{Footprint of the SPIRE FTS detectors (corresponding to the observation obsid:1342214844) over the SPIRE 250 $\mu$m image centered at the position of Sgr B2(M). Sgr B2(M) is observed with the central detectors SLWC3-SSWD4 and Sgr B2(N) with the detectors SLWC2 and SSWF3. The white circle represents the FoV with a diameter $\sim$2 arcmin.}
\label{fig2}
\end{figure}

	\par{In this paper we study the physical and chemical conditions in the Sgr B2 molecular cloud by analyzing the entire submm band from 447 GHz to 1550 GHz spectra taken with the {\it Herschel} SPIRE Fourier Transform Spectrometer (FTS). 
}

	\par{The SPIRE FTS data allow us to study the molecular gas by modeling the CO rotational line emission using a non-LTE radiative transfer model and to make the maps of several molecular and atomic lines across the region. We also analyze the large-scale far-infrared (far-IR) dust emission from Sgr B2 using {\it Herschel} PACS and SPIRE observations that were obtained as part of the Hi-GAL key program \citep{Molinari10}. 
}
\begin{figure*}[t!]
        \centering
        \begin{subfigure}[]
                \centering
 \includegraphics[width=6.cm]{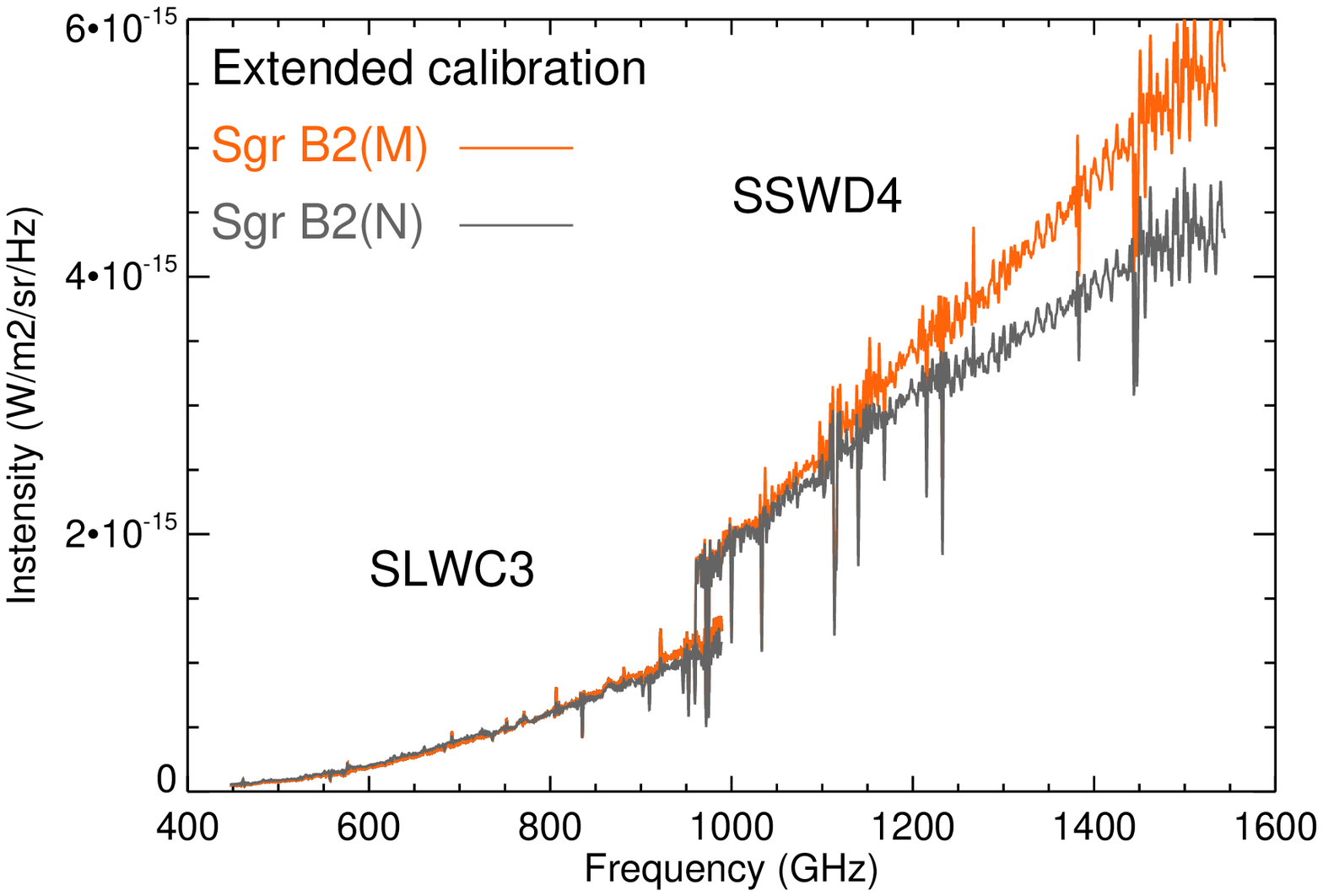}
                       \end{subfigure}
        \begin{subfigure}[]
                \centering
  \includegraphics[width=6.cm]{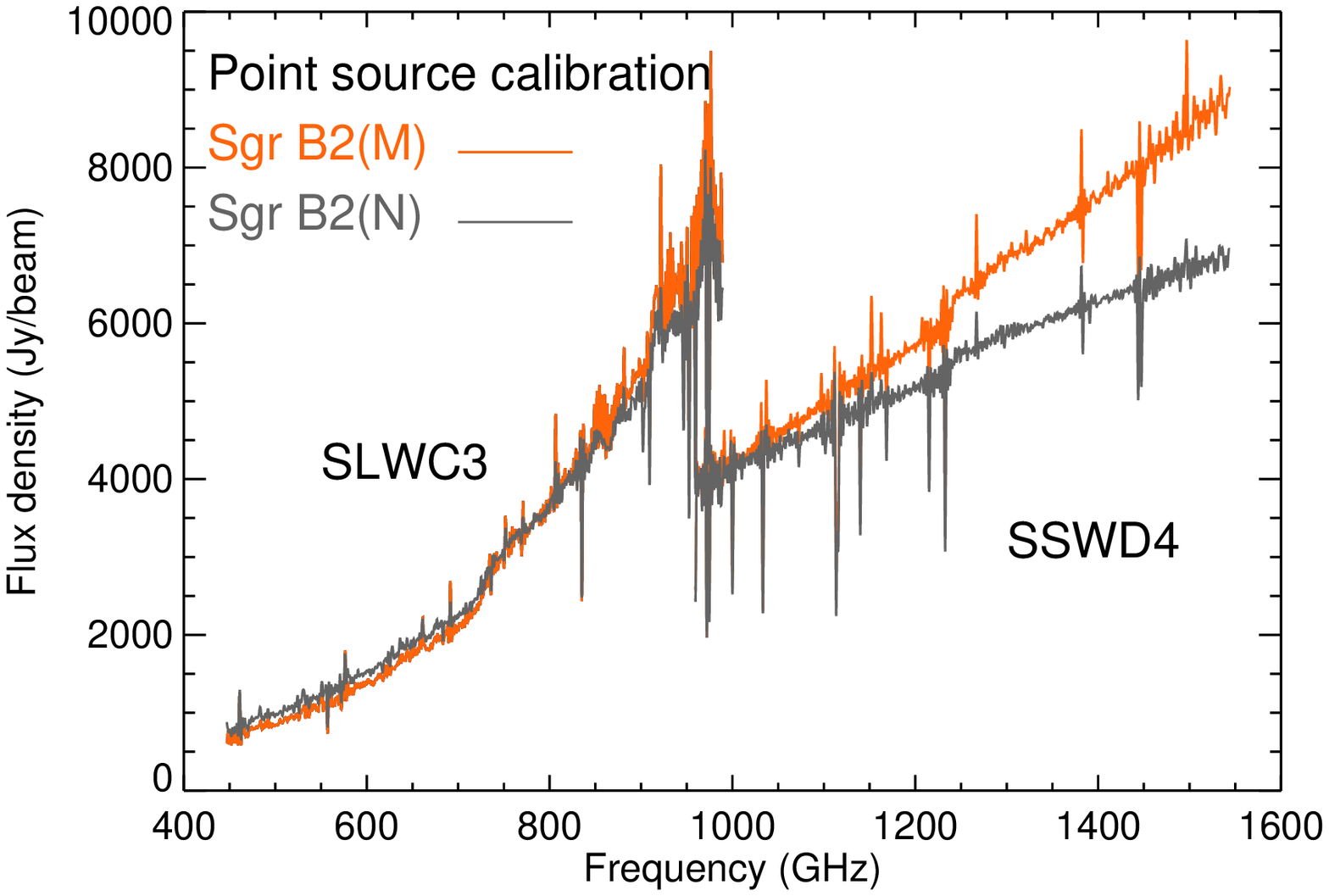}
                        \end{subfigure}
        \begin{subfigure}[]
                \centering
   \includegraphics[width=6.cm]{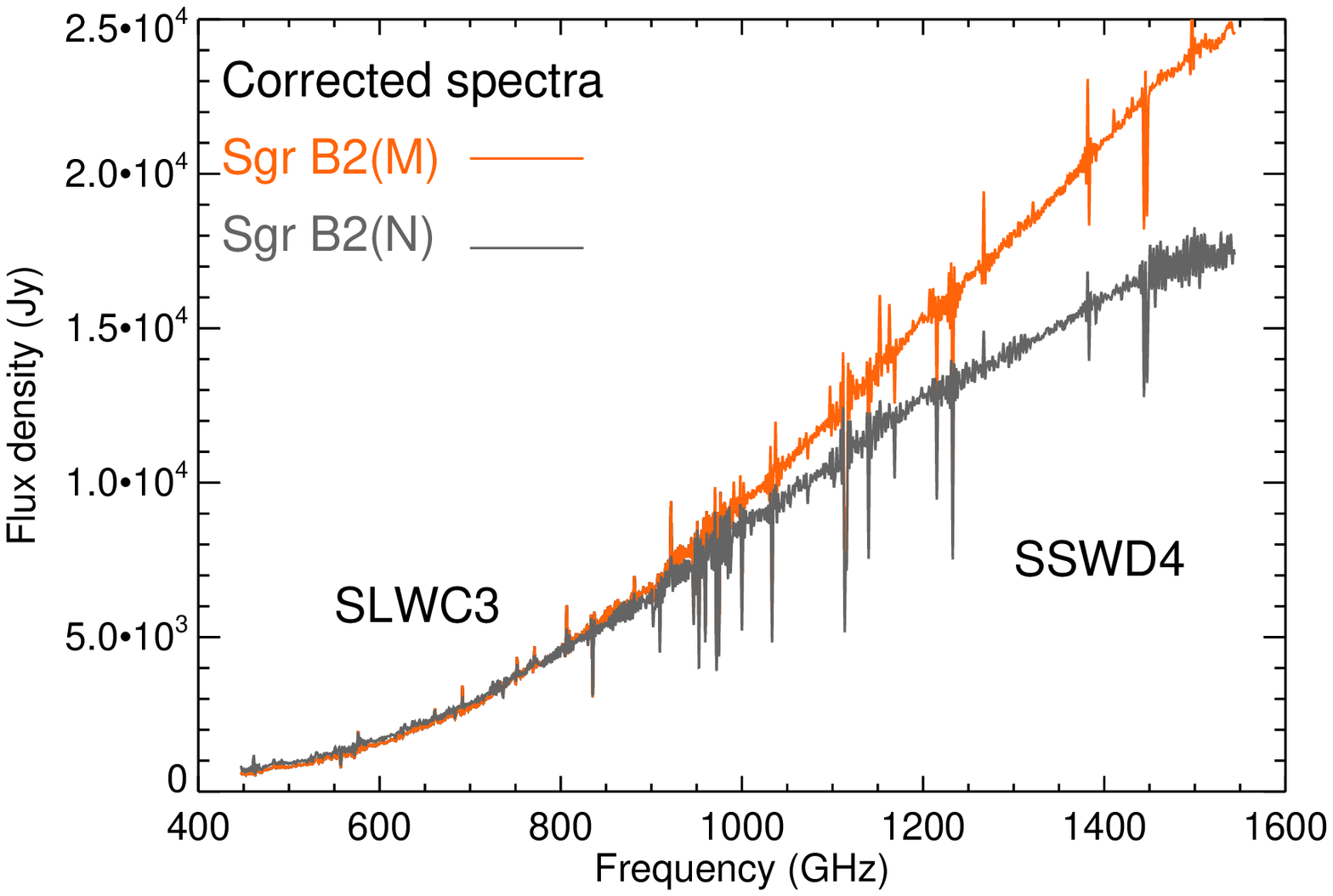}
                      \end{subfigure}
        \caption{(a) SPIRE FTS unapodized spectra at the positions of Sgr B2(M) (orange) and Sgr B2(N) (grey) obtained with the extended calibration. (b) Same spectra in Jy/beam (with variable beam size in frequency) obtained with the pointed calibration mode. (c) Final spectra in Jy (within a beam size of FWHM= 40"), corrected with the new calibration tool developed for semi-extended sources.}\label{fig3}
\end{figure*}
\section{Observations and Data Reduction}\label{Obser}
  \subsection{Spectroscopic Data}\label{spec-data}        
    \subsubsection{SPIRE FTS}\label{spire-fts}
	\indent\par{The SPIRE FTS \citep{Griffin10}, onboard the {\it Herschel} space observatory \citep{Pilbratt10}, observed Sgr B2 as part of the SPECHIS GT proposal: ``SPIRE Spectral Line Survey of HIFI-GT-KP-Sources program'' (PI. E.~Polehampton) with two bolometer arrays, the SPIRE Short Wavelength Spectrometer Array (SSW) with 37 detectors covering the wavelength range $194-313$ $\mu$m and the SPIRE Long Wavelength Spectrometer Array (SLW) with 19 detectors covering the range $303-671$ $\mu$m, with a pixel spacing of approximately twice the beam. Seven SLW detectors spatially sample the field of view (FoV), which is $\sim$2 arcmin in diameter. Two single pointing observations were made at high resolution (0.04 cm$^{-1}$) in bright mode to avoid saturation: ObsId:1342214843, centered at the position of Sgr B2(N) (RA(J2000)= 17$^{\rm h}$47$^{\rm m}$20.00$^{\rm s}$ Dec.(J2000)= -28$^{\rm o}$22'17.44''), and ObsId:1342214844, centered at the position of Sgr B2(M) (RA(J2000)= 17$^{\rm h}$47$^{\rm m}$20.30$^{\rm s}$ Dec(J2000)= -28$^{\rm o}$23'04.01''). The total integration time was 266 s for each observation and the radial velocity of the telescope with respect to the local standard of rest (LSR) was ${\rm v_{\rm rad}}=$ 39.32 km$\ $s$^{-1}$ at the moment of the observation. The full width half maximum (FWHM) of the SPIRE FTS spectra is wavelength dependent, changing between 17" and 21" for the SSW band and between 37" to 42" for the SLW band \citep{Griffin10}.
}
	\par{Figures~\ref{fig1} and~\ref{fig2} show the layout of the spectrometer detector arrays, SLW (black) and SSW (blue), for each observation covering an area of 3.5 arcmin in diameter over the PACS 70 $\mu$m photometric image. The central detectors SLWC3 and SSWD4, are pointing at the position of Sgr B2(N) in Figure~\ref{fig1} and at the position of Sgr B2(M) in Figure~\ref{fig2}. There are two dead detectors in the SSW array: SSWF4 and SSWD5.
 }
	\par{The SPIRE FTS data were reduced with the {\it Herschel} Interactive Processing Environment (HIPE) version 9.2. The  SPIRE FTS reduction pipeline provides data calibrated with two assumptions about the source size: an extended emission over a region much larger than the beam or truly point-like emission centered on the optical axis. Details on the calibration of the SPIRE FTS data are given in \citet{Wu12}. However, neither of these assumptions fits with the Sgr B2 cores, which are semi-extended source. If the point-source calibration or the extended calibration is applied to a semi-extended source such as Sgr B2(M) or Sgr B2(N), a discontinuity appears in the overlap wavelength range of the continuum measured by the SSW and the SLW detectors. This is due to the variation of the beam diameter from one detector to the other (see Figure~\ref{fig3}). In this work, the discontinuity in the SPIRE FTS spectra was corrected with a new calibration tool for semi-extended sources based on a model of the source distribution and the beam profile shape. The tool is based on the point source calibration, and a detailed explanation of this correction tool can be found in \citet{Wu12}.
}


	\par{Figure~\ref{fig3} shows the unapodized spectra at the positions of Sgr B2(M) and Sgr B2(N) calibrated as an extended source (left), as a point source (center), and after the application of the new calibration tool (right) on the spectrum previously calibrated as a point source (center). The best continuum matching for both spectral continuum levels is obtained by assuming a Gaussian emission profile for each source, with FWHM= 30" and FHWM= 23" (these are equivalent to a radius  of $\sim$ 0.62 pc and $\sim$ 0.47 pc for Sgr B2(M) and Sgr B2(N), respectively).
}
	\par{The angular resolution of the resulting spectrum after the beam size correction is that of the SLW detector in the overlapping range of frequencies, which corresponds to FWHM= 40"; this is equivalent to a radius of $\sim$ 0.8 pc.
} 
\subsubsection{Line Mapping with the SPIRE FTS}\label{Obs-map}
	\indent\par{The integrated line intensities were obtained from the unapodized SPIRE FTS spectra after baseline subtraction. To get a good accuracy on the position of the central frequency of spectral lines, the correction of the frequency scale was done by converting it to the LSR frame, taking into account the radial velocity (${\rm v_{\rm rad}}$) of the {\it Herschel} telescope along the line of sight, so that the corrected central frequency was calculated as $\nu_{\rm corr.}=\nu (1- {\rm v_{\rm rad}}/{\rm c})$. The line shape was fitted with a classical $sinc$ function using the line-fitting tool available in HIPE version 9.2. Sparse-sampling maps for several emission and absorption lines were made by interpolating the value of the integrated intensity of each line measured by adjacent detectors to halfway positions in order to analyze the spatial distribution of these species. 
}
	\par{The integrated intensities of the selected lines were obtained for the spectra observed by each detector based on the extended source calibration in order to get maximum spatial 
coverage. This was done without correcting the discontinuity in the overlap region of the continuum with the new correction tool because, as explained in the previous subsection, the tool was developed for point source calibration, which is based on Uranus observations. However, Uranus was observed with only the 7 SLW and the 17 SSW detectors inside the field of view. The point source calibration was not carried out for the outer detectors, and the correction of the continuum with the tool would imply a reduction of the spatial coverage of the maps of the molecular and ionic lines. Therefore, the integrated line intensities depend on the beam size for each frequency; hence each map presents a different beam size.
}
	\par{The total area of each map is $\sim 3.5\times 3.5$ arcmin$^2$. The maps of those lines observed with the SLW detector have an angular resolution FWHM$\sim 30-35$". The maps of the lines in the frequency range of the SSW detector are spatially better sampled with an angular resolution FWHM$\sim 18$". All maps in Figure~\ref{fig12} are centered at the position of Sgr B2(M) (RA= 17$^{\rm h}$47$^{\rm m}$20.40$^{\rm s}$, Dec= -28$^{\rm o}$23'03.84").
} 
\subsubsection{PACS}\label{obs-pacs}
\begin{figure}[t]
\resizebox{\hsize}{!}{\includegraphics{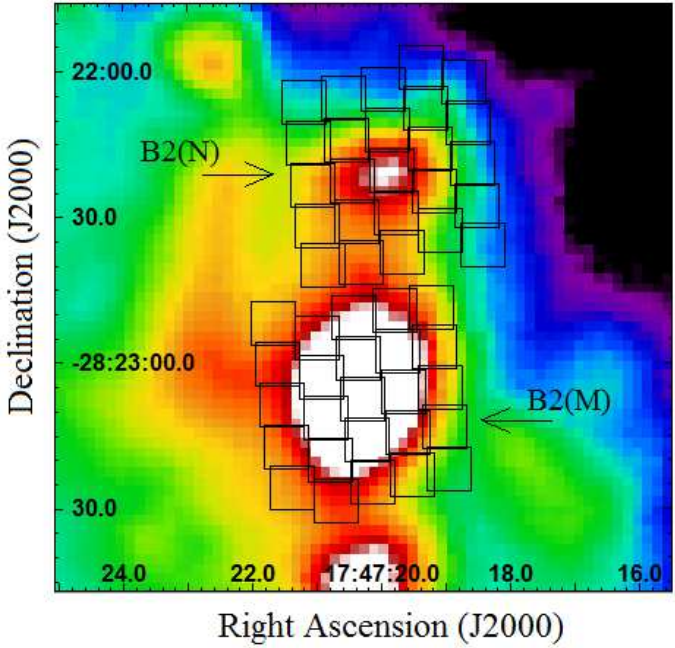}}
\caption{Two footprints of the PACS spaxels grid overlaid on the PACS 70 $\mu$m image and centered at the position of Sgr B2(M) and Sgr B2(N). Each grid covers $\sim$47$\times$47 arsec$^2$ ($\sim$2$\times$2 pc$^2$).}
\label{fig4}
\end{figure}
	\indent\par{To complement the mid-J CO lines observed with SPIRE FTS, we also present the first detection of the $^{12}$CO $J$=14-13 (186.13 $\mu$m), 15-14 (173.75 $\mu$m), and 16-15 (162.92 $\mu$m) lines towards Sgr\,B2 (M) and (N) with the PACS spectrometer \citep{Poglitsch10}. These data are part of the HEXOS Guaranteed-Time Key-Program (PI. E.~Bergin).
}
	\par{The PACS spectrometer uses photoconductor detectors and provides 25 spectra over a 47$''$$\times$47$''$ FoV resolved in 5$\times$5 spatial pixels (``spaxels''), each with a size of $\sim$9.4$''$ on the sky (Figure~\ref{fig4}). The resolving power is $\lambda$/$\Delta$$\lambda$$\sim$1000-1500 in the R1 grating order where CO lines are detected. }

	\par{The observations were carried out in October 2010 and April 2011 in the range spectroscopy unchopped mode, using a distant position as reference position (we checked that no extended CO emission appears in the OFF positions). The corresponding observation ObsID are 1342206883 (ON centered at Sgr\,B2(M) core), 1342206881 (OFF reference position), 1342218190 (ON centered at Sgr\,B2(N) core), and 1342218188 and 1342218192 (OFF reference positions).
}
	\par{The Sgr\,B2 sources are very bright in the far-IR (e.g., above 6000\,Jy at $\sim$120\,$\mu$m in a PACS spaxel towards Sgr B2(M)). Hence, in order to avoid saturation, these observations were observed in a nonstandard engineering
mode.  The complete spectra and detailed data reduction technique will be presented in Godard et al.~(2013, in preparation). The measured width of the spectrometer point spread function (PSF)  is relatively constant for $\lambda$$\lesssim$100\,$\mu$m (FWHM$\simeq$ spaxel size) but increases at longer wavelengths.
In particular, only $\sim$40\%\, of the emission from a point source would fall in the central spaxel at $\gtrsim$160 $\mu$m, preventing the production of the CO integrated intensity maps.  Since Sgr B2 cores are extended sources in comparison to the spaxel size, the CO line fluxes were extracted by adding the observed flux of all the 5$\times$5 spaxels.
}
\subsection{Photometric Data}\label{Phot-data}
\subsubsection{PACS and SPIRE}\label{pacs-spire}
\begin{figure}[t]
\resizebox{\hsize}{!}{\includegraphics{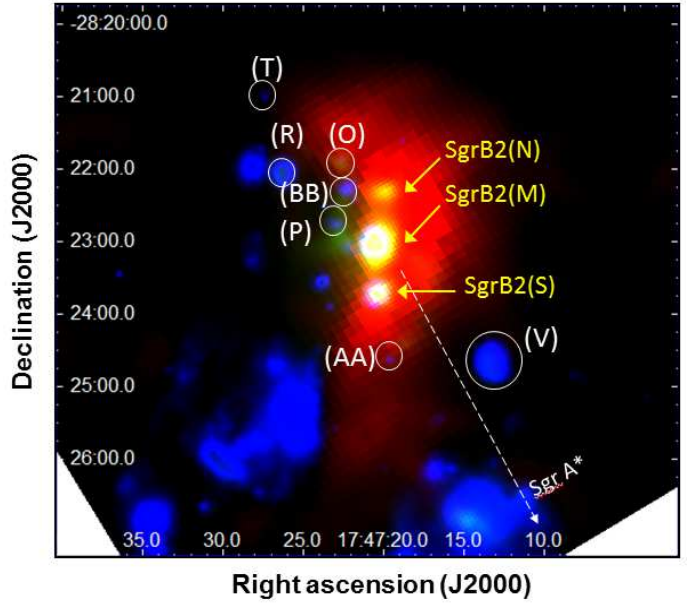}}
\caption{Composite image of the Sgr B2 molecular cloud with the MIPS 24 $\mu$m (blue), PACS 70 $\mu$m (green), and SPIRE 350 $\mu$m (red). The image shows the position of the main three star-forming cores and the main compact and ulta-compact H{\sc ii} regions in the complex. The white-dashed arrow points out the direction of Sgr A$^*$ at the Galactic Center.}
\label{fig5}
\end{figure}

	\indent\par{The Sgr~B2 molecular cloud was observed with the PACS and the SPIRE photometers as part of the Hi-GAL Key Project \citep{Molinari10}. The processed image of Sgr B2 is part of the $2^{\rm o}\times 2^{\rm o}$ region at the Galactic Center, mapped simultaneously in parallel mode with the PACS and the SPIRE photometers \citep{Molinari11}. The basic data reduction procedures are given in \citet{Traficante11}. Fully sampled images were made at five different wavelength bands: 70 $\mu$m, 160 $\mu$m, 250 $\mu$m, 350 $\mu$m, and 500 $\mu$m, with FWHM= 5.9", 11.6", 18.5", 25.5", and 36.8", respectively. Images are made in MJy/sr at 3.2"/pixel and 4.5"/pixel for PACS 70 $\mu$m and PACS 160 $\mu$m, respectively, and at 6.0"/pixel, 8.0"/pixel, and 11.5"/pixel for SPIRE 250 $\mu$m, SPIRE 350 $\mu$m, and SPIRE 500 $\mu$m, respectively. 
} 
	\par{To avoid saturation and minimize the non-linearity, SPIRE was used in ``bright-source mode". The PACS and SPIRE photometric data were used to determine the dust temperature, hydrogen column density, and dust spectral index distribution across the Sgr B2 molecular cloud. The PACS 160 $\mu$m image was discarded in the calculations due to the saturation effects at the position of the main cores. The spectral energy distribution (SED) for each pixel in the {\it Herschel} image was fitted after smoothing the images and adjusting the pixel sizes uniformly to the resolution of the SPIRE 350 $\mu$m channel band (25.5'').
}
\subsubsection{Spitzer MIPS 24 $\mu$m}\label{mips}


	\par{The Spitzer Space Telescope observed the inner Galactic plane using the multiband infrared photometer for Spitzer (MIPS) in two passbands, 24 and 70 $\mu$m, as part of the Spitzer MIPSGAL Legacy Survey \citep{Carey09}. Although the innermost regions of the Galactic Center are saturated in MIPS 24 $\mu$m, there are only a couple of saturated pixels at the cores of the brightest compact sources in the data covered by the SPIRE FTS observations of SgrB2. The values of these pixels have been replaced with those at 21.3 $\mu$m from the Midcourse Space Experiment (MSX) \citep{Price01} and color corrected for a star-forming region emission. The surface brightness of the pixels closest to the affected ones within this radius can change by as much as 30 \%, but this is a tiny area around the core. We used Spitzer 24 $\mu$m to trace the hot dust and the very small grains (VSG) population across the Sgr B2 complex.
}

\section{Results}\label{results}

\subsection{Dust-Extended Emission}\label{dust}


\indent\par{Figure~\ref{fig5} shows the Sgr B2 molecular complex as seen by combining the images from MIPS at 24 $\mu$m (blue), PACS 70 $\mu$m (green), and SPIRE 350 $\mu$m (red). The size of the image is $\sim 10\times10$ arcmin$^2$ ($\sim 24.75\times24.75$ pc$^2$), and it is centered at RA(J2000)=17$^{\rm h}$47$^{\rm m}$21.01$^{\rm s}$ Dec(J2000)= -28$^{\rm o}$23'06.28", close to the center of Sgr B2(M). The white-dashed arrow points out to the direction of Sgr A$^*$ at the Galactic Center.
}
\par{The PACS and the SPIRE observations provide a view of the three main star-forming cores and more extended structures in the Sgr B2 region with the highest resolution and best sensitivity to date in the far-infrared wavelength range, which allowed us to study the dust properties' distribution in the entire region and resolve the cores from the extended envelope.
}

\par{{\it Herschel} PACS and SPIRE cover the wavelength range between 70 $\mu$m and 500 $\mu$m. At these wavelengths ($\lambda>$ 60 $\mu$m), the SED is dominated by thermal dust emission due to the population of big grains (BGs), peaking at $\sim$100 $\mu$m. The PACS 70 $\mu$m emission traces the warm dust, which is mainly distributed through the three main compact cores: Sgr B2(N), Sgr B2(M), and Sgr B2(S). These sources are known to be  active sites of high-mass star-formation. 
}

\par{The SPIRE 350 $\mu$m image traces the distribution of colder dust. The map shows an extended cloud mainly distributed through the main star-forming cores and spreading out towards the north, west and south. 
}


\par{Sgr B2 also contains several compact and ultracompact H {\sc ii} regions dominated by free-free emission at wavelengths $\lambda > 3$ mm, while at shorter wavelengths thermal emission from dust dominates \citep{Gordon93}. The MIPS 24 $\mu$m image shows these compact H {\sc ii} regions mainly distributed at the east of the three main star-forming cores. The MIPS map at 24 $\mu$m can be used to trace the population of VSGs. With sizes of $\sim$0.01 $\mu$m, the VSGs can increase their temperature up to 80 K, absorbing a single UV photon, and re-emit at shorter wavelengths ($\lambda\sim$ 24 $\mu$m), so that MIPS 24 $\mu$m traces the hottest dust \citep{Desert90}. MIPS 24 $\mu$m shows the population of VSGs distributed mainly through the compact H {\sc ii} regions Sgr B2(R), Sgr B2(V), and Sgr B2(AA), as well as several ultracompact H {\sc ii} regions (Sgr B2(O), Sgr B2(T), Sgr B2(P), and Sgr B2(BB)). These regions are distributed from north to south, along the east side of the three main cores. Observations of these sources at radio wavelengths reveal details that are not possible to resolve at the resolution of MIPS 24 $\mu$m. The ultracompact H{\,\sc ii} regions B2(R) and B2(U) seem to be an edge-brightened shell-like source \citep{Mehringer93}. Region B2(V) is also a shell-like source with a clumpy structure \citep{Mehringer93}. These compact H {\sc ii} regions are hardly visible in SPIRE 350 $\mu$m data. Sgr B2(M) and Sgr B2(S) are also bright on the MIPS 24 $\mu$m map; however, Sgr B2(N) is unresolved at 24 $\mu$m.  
}

\begin{figure}[t]
\resizebox{\hsize}{!}{\includegraphics{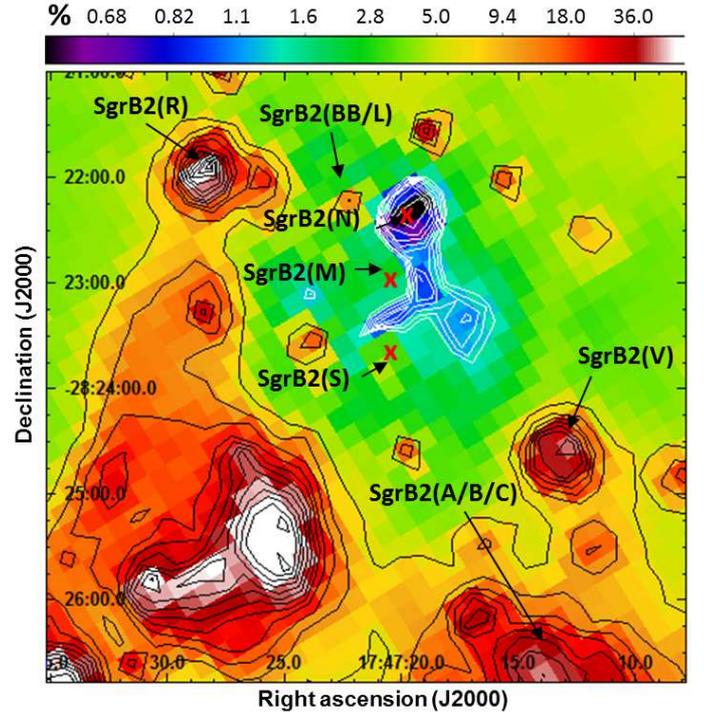}}
\caption{Plot represents an upper limit to the total contribution of the VSGs emission to the dust thermal radiation at 70 $\mu$m calculated as $f_{24\mu m}/f_{70 \mu m}\times 100$. The white contours represent the area with the lowest contribution ($<1.8\%$). Black contours trace the regions with the highest contributions ($>8\%$).} 
\label{fig6}
\end{figure}


\subsubsection{Very Small Grain Contribution to the PACS 70 $\mu$m intensity}\label{vsg}


\indent\par{The SED of the VSGs peaks at $\sim$24 $\mu$m and extends to longer wavelengths, contaminating the dust thermal emission at $\lambda<$ 100 $\mu$m. This produces an excess in the PACS 70 $\mu$m intensity and can therefore cause an increase in the estimation of the dust temperature. The contribution of the emission of the VSGs to the PACS 70 $\mu$m intensity is expected to be more significant in those regions where the UV radiation is stronger and the abundance of VSGs is larger (i.e., near YSOs, H {\sc ii} regions, UV radiation illuminated cloud surfaces, etc).
}
\begin{figure*}[!t]
        \centering
        \begin{subfigure}[]
                \centering
 \includegraphics[width=6.cm]{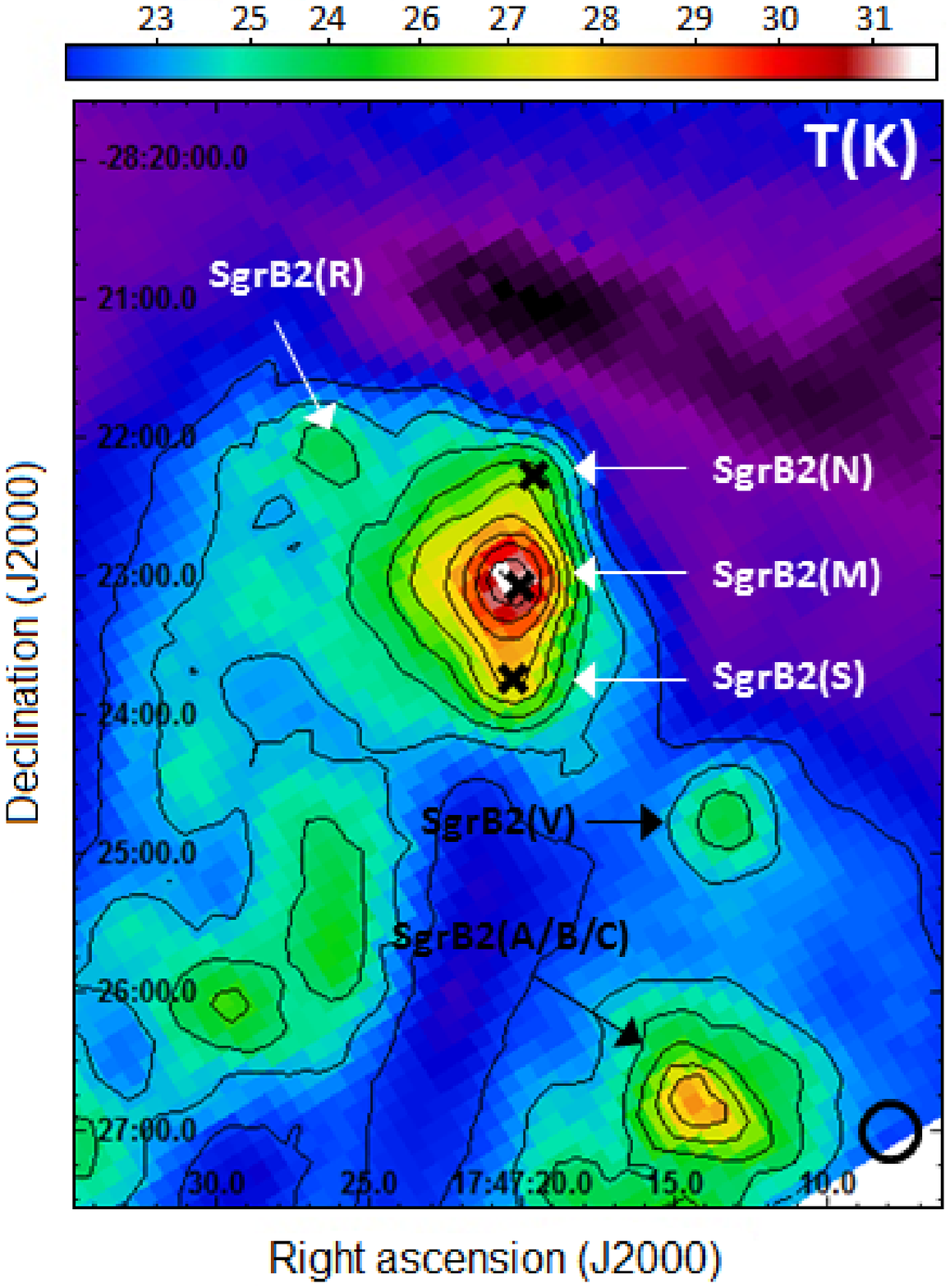}
              \end{subfigure}
        \begin{subfigure}[]
                \centering
  \includegraphics[width=6.cm]{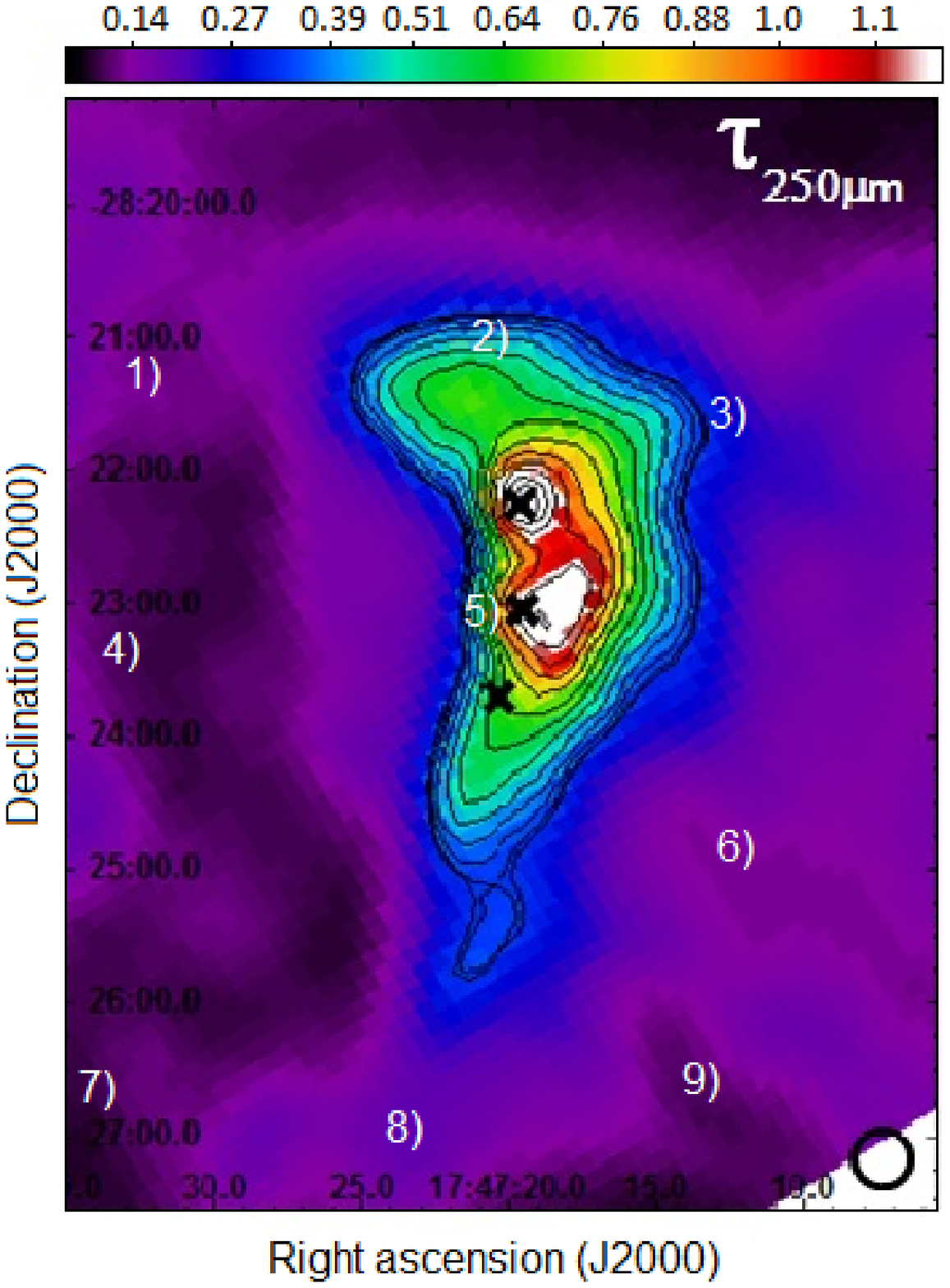}
           \end{subfigure}
        \begin{subfigure}[]
                \centering
   \includegraphics[width=6.cm]{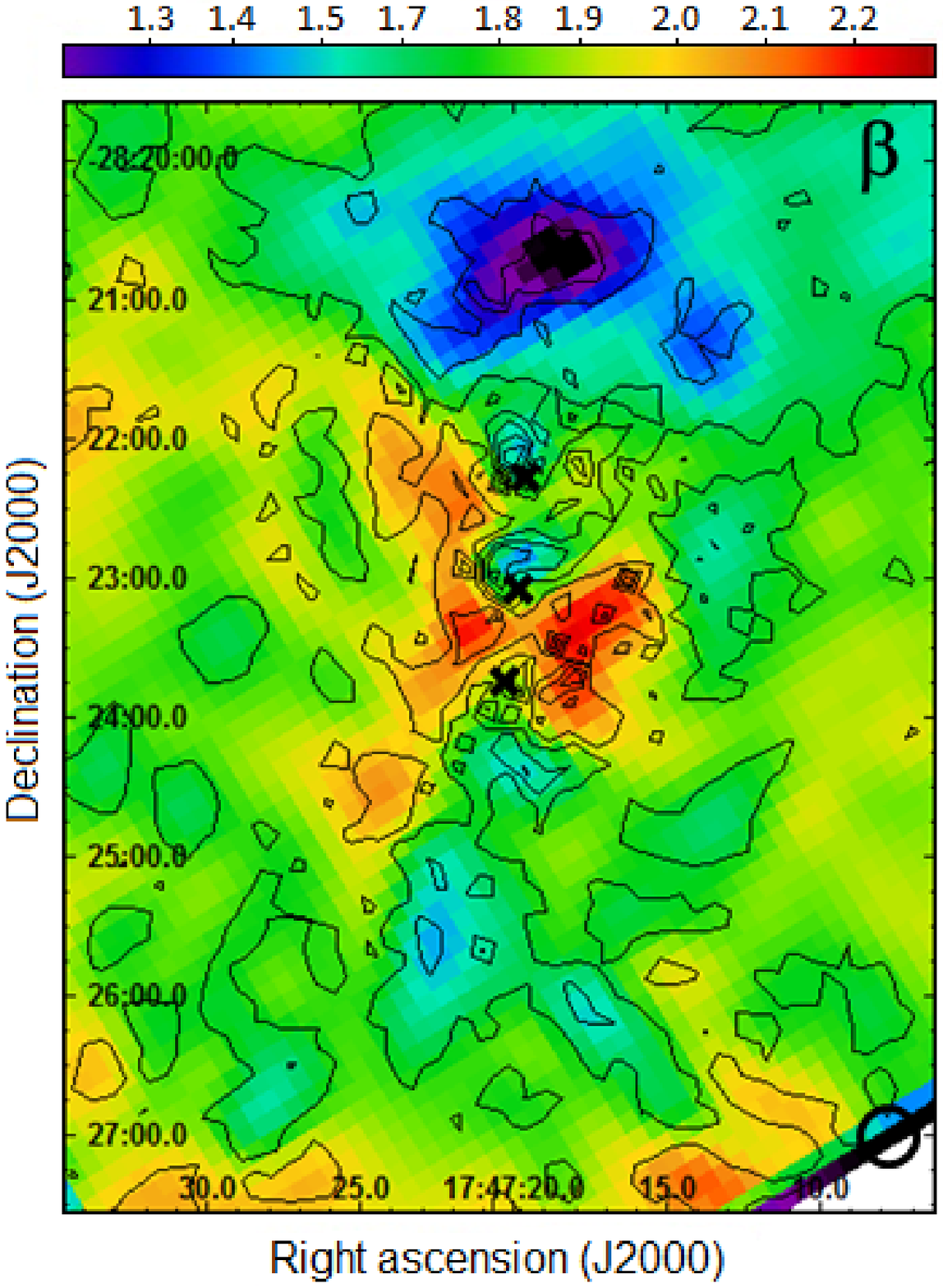}
               \end{subfigure}
       \caption{(a) Dust temperature map of the Sgr B2 molecular cloud. (b) Optical depth image at 250 $\mu$m. (c) Map of the dust emissivity index, $\beta$ (see Eq.~\ref{eq:tau}). The size of the beam (FWHM= 36.9") is plotted at the down-right side of the images. The size of the images is $\sim$7.0$\times$ 8.22 arcmin$^2$, equivalent to $\sim$17.3$\times$20.3 pc$^2$. White numbers on the optical depth image represent the nine positions, whose SED and fitting are shown in Figure~\ref{fig8}.} 
\label{fig7}
\end{figure*}
\par{The MIPS 24 $\mu$m map was used to obtain a qualitative idea of the contribution of the VSGs' emission to the total dust emission at 70 $\mu$m along Sgr B2: $f_{24\mu m}/f_{70 \mu m}\times 100$. This represents an upper limit to the total contribution since the intensity of the VSGs' emission at 70 $\mu$m is lower than the intensity at 24 $\mu$m (which is approximately the peak of the VSGs' emission). The distribution of this contribution along Sgr B2 is shown in Figure~\ref{fig6}. In those regions where $f_{24\mu m}/f_{70 \mu m}\times 100$ is small, the emission from VSGs is less significant at 70 $\mu$m and therefore a substantial part of dust emission is related to warm BGs. The contribution of the VSGs' emission to the thermal emission at 70 $\mu$m can be as high as $\sim$ 60$\%$ along the Sgr B2(R) and $\sim$ 36$\%$ through Sgr B2(V) and Sgr B2(A/B/C) H {\,\sc ii} regions. The position with the lowest contribution of VSG emission is at the position of Sgr B2(N) with a contamination $\leq$0.6 $\%$. At the west side of Sgr B2(M) and Sgr B2(S) and at the region between Sgr B2(M) and Sgr B2(S), the contamination is $<1.8\%$. These two sources present a contribution of VSG emission of 2$\%$ and 5$\%$ to the total, respectively. The second peak with the lowest contribution is located at a distance of $\sim 0.4$ arcmin to the southwest of Sgr B2(M). 
}

\subsubsection{Dust Temperature Distribution}\label{T-dust} 
\indent\par{Since the contribution of VSG emission to the far-infrared thermal dust emission at 70 $\mu$m is almost negligible around the main three star-forming cores, PACS 70 $\mu$m, together with the SPIRE images at 250 $\mu$m, 350 $\mu$m, and 500 $\mu$m, can be used to calculate the dust temperature, the spectral index, and the hydrogen column density through the region.
}

\par{The dust continuum emission can be fit by a modified black body curve, which depends on the dust temperature, the spectral index and the optical depth at different frequencies:
}

\begin{equation}
F_{\nu}=\Omega\times B_{\nu}\left( T_d \right)\times\left( 1-e^{-\tau_{\nu}} \right), 
\label{eq:greybody}
\end{equation}

\noindent where $\Omega$ is the beam solid angle, $T_{\rm d}$ is the dust temperature, and $\tau_{\nu}$ is the optical depth for a frequency $\nu$, given as

\begin{equation}
\tau_{\nu}=\tau_0 \left(\frac{\nu}{\nu_0}\right)^{\beta}, 
\label{eq:tau}
\end{equation}  
\noindent where $\tau_0$ is the reference optical depth at a reference frequency $\nu_0$ corresponding to 250 $\mu$m. 


\par
\begin{figure}[!t]
\centering
\resizebox{\hsize}{!}{\includegraphics{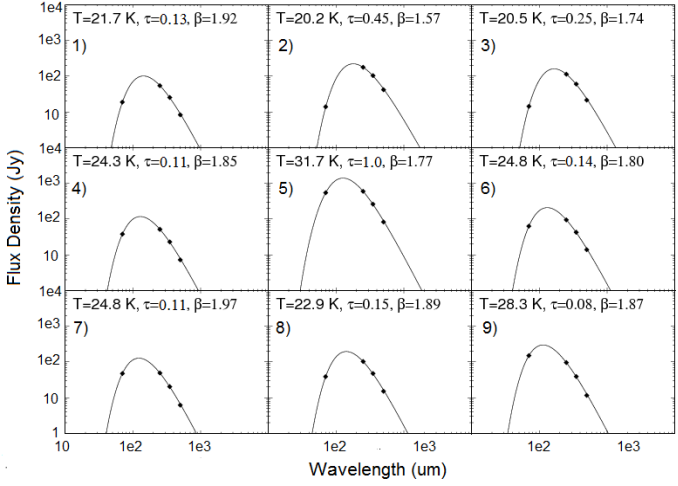}}
\caption{SED fitted by a modified black body curve in nine different positions. Each position is shown in the optical depth image in Figure~\ref{fig7}. Fluxes are smoothed to the resolution of SPIRE 500 $\mu$m (FWHM= 36.9").} 
\label{fig8}
\end{figure}

{We have used the IDL code MPFITFUN based on a least-square $\chi^2$ SED fit to produce the T, $\beta$ and $\tau_{\nu}$ images of Sgr B2. Figure~\ref{fig8} shows the SED fitted by a modified blackbody curve and the obtained values of T, $\beta$, and $\tau_{\nu}$ at nine different positions. Figure~\ref{fig7}a shows the dust temperature distribution through the Sgr B2 molecular cloud. The maximum value of the dust temperature is located just at the position of Sgr B2(M), with values $T_{\rm d}\sim$ 34 K. Sgr B2(S) and Sgr B2(N) are slightly colder, with $T_{\rm d}\sim$ 30 K and $T_{\rm d}\sim$ 28 K, respectively. The temperature distribution extends from the center of the image, on Sgr B2(M), to the southeast and to the southwest, with values  $T_{\rm d}\sim$ 25 K.  The temperature contours also trace the H {\sc ii} region Sgr B2(A/B/C) with $T_{\rm d}\sim$ 26 K and Sgr B2(V) with temperature values of $T_{\rm d}\sim$ 25 K. The coldest region is placed 1.2 arcmin at the north of Sgr B2(N) with values $T_{\rm d}\sim$ 20 K and extends to the west with temperatures lower than $\sim$ 21 K.
}
\subsubsection{Optical Depth}\label{T-optdepth}


\indent\par{Figure~\ref{fig7}b shows the optical depth image calculated at 250 $\mu$m. The optical depth contours trace the regions with the largest column density of material on the image extending from north to south through the center of the mapped region, with the largest values peaking at the position of Sgr B2(N) ($\tau_{250}\sim$ 1.6). A second peak is observed on the image with a slightly lower optical depth value ($\tau_{250}\sim$ 1.4). The peak is located just at the southwest side of Sgr B2(M), coinciding with the region of low VSG emission contribution, which is located at a distance of $\sim$0.4arcmin at the southwest of Sgr B2(M) (Figure~\ref{fig6}). Sgr B2(M), however, shows lower optical depth values than this region  ($\tau_{250}\sim$ 0.9), and the optical depth in Sgr B2(S) is $\tau_{250}\sim$ 0.6. The optical depth can be approximately converted into a total gas column density ($N{\rm (H)}= N{\rm (H\,{\sc I})}+2N{\rm (H_2)}$) by { $\tau_{250} = 8.8\times 10^{-26}N{\rm (H)}$ cm$^2$/H} \citep{Bernard10}.
}
\begin{figure}[!t]
\centering
\resizebox{\hsize}{!}{\includegraphics{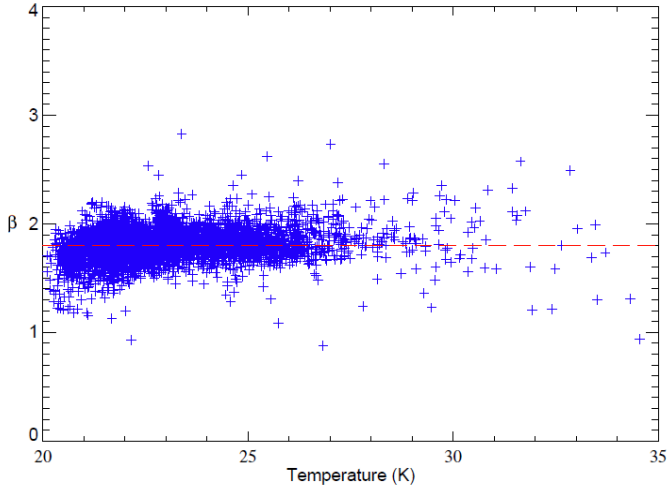}}
\caption{Distribution of dust temperature versus the spectral index ($\beta$) in Sgr B2. The dashed red line represents the mean value of $\beta$, which is $\sim$1.8.} 
\label{fig9}
\end{figure}

\begin{table*}[!ht]
\centering
\caption{Observed line surface brightness obtained from the SPIRE FTS spectra of Sgr B2(M), Sgr B2(N), and of the extended envelope at an offset $(x',y')=(-1.24',0')$ from Sgr B2(M). $\Omega_{\rm beam}(\nu)$ represents the beam solid angle for the extended source calibration.} 
\label{tab1}
\begin{tabular}{ccccccc}
\hline\hline
 {Species} & {Transition}  & {$\nu$} &{$\Omega_{\rm beam}(\nu)$ }&{Sgr B2(M)} &{Sgr B2(N)} &{Extended envelope}\\
   &   & {(GHz)}&(sr$\ 10^{-8}$)&{(W$\ $m$^{-2}$$\ $sr$^{-1}$$\ 10^{-8}$)} &{(W$\ $m$^{-2}$$\ $sr$^{-1}$$\ 10^{-8}$)}&{(W$\ $m$^{-2}$$\ $sr$^{-1}$$\ 10^{-8}$)}\\
\hline
{[C\,{\sc i}]} & $J=1-0$ & 492.06 & 12.6  &2.1$\pm$0.4 &1.8$\pm$0.2&1.1$\pm$0.2 \\
{[C\,{\sc i}]} & $J=2-1$ & 809.41 &  8.62  &5.5$\pm$0.6 & 4.0$\pm$0.7& 5.0$\pm$0.6 \\
CH$^+$ & $J=1-0$ & 835.45& 8.89  &-36.1$\pm$7.4 & -33.6$\pm$4.7&-7.0$\pm$0.9 \\
CO     & $J=4-3$     & 461.02 & 14.2   &4.6$\pm0.9$ &3.8$\pm$0.6 &  5.1$\pm$0.6\\
CO     & $J=5-4$     & 576.39 &  9.54  &9.2$\pm$2.01  & 5.5$\pm$0.7& 6.8$\pm$0.7 \\
CO     & $J=6-5$     & 691.51 & 8.06   &18$\pm$3.9& 7.2$\pm$1.2& 8.9$\pm$1.3\\
CO     & $J=7-6$     & 806.65 & 8.59   &22$\pm$4 & 8.9$\pm$1.3 &11.3$\pm$1.3\\
CO     & $J=8-7$     & 921.77 & 9.92   &34$\pm$7 & 15.1$\pm$2.0& 10.7$\pm$1.6\\
CO &$J=9-8$&1036.88 & 2.50    &40$\pm$8& 20$\pm$6& 6.4$\pm$1.9\\
CO & $J=10-9$& 1152.47 &  2.17     &53$\pm$9 & 27$\pm$7&4.2$\pm$1.0 \\
CO &$J=11-10$ & 1267.09 &   2.08   &60$\pm$10 & 26.0$\pm$8.4& 1.6$\pm$0.5\\
CO &$J=12-11$ & 1381.99 &  2.13    &52$\pm$10 & 30.1$\pm$7.7& 0.8$\pm$0.2\\
CO$^{\mathrm{\star}}$ &$J=14-13$ & 1611.8 & 2.34    &30$\pm$6 & 16$\pm$3& ...\\
CO$^{\mathrm{\star}}$  & $J=15-14$&1726.6&   2.33   &20$\pm$5 & 12$\pm$3& ...\\
CO$^{\mathrm{\star}}$  & $J=16-15$&1841.3&   2.14   &13$\pm$3& 7.6$\pm$1.6 & ...\\
$^{13}$CO& $J=5-4$& 551.11&  10.3   &4.0$\pm$0.8 & 2.4$\pm$0.3& 0.7$\pm$0.1\\
$^{13}$CO& $J=6-5$ & 661.20 &  8.20   &6.2$\pm$0.9& 3.6$\pm$0.7& 1.1$\pm$0.2\\
$^{13}$CO & $J=7-6$& 771.19 &   8.28   &9.7$\pm$1.8& 5.5$\pm$1.1& 1.0$\pm$0.2\\
$^{13}$CO & $J=8-7$& 881.23 &  9.43     &12.5$\pm$2.6 &6.4$\pm$0.7 & 0.8$\pm$0.1\\
$^{13}$CO&$J=9-8$& 991.50 &   2.71   &15$\pm$4 &7.3$\pm$2.3 & ...\\
$^{13}$CO&$J=10-9$& 1100.62&  2.28  &8.6$\pm$2.8  &9.2$\pm$2.4 & ...\\
$^{13}$CO&$J=11-10$& 1211.00 &   2.10      &35.8$\pm$9.6 &22.6$\pm$6.3& ...\\
$^{13}$CO&$J=12-11$ & 1321.40& 2.09      &11$\pm$3 &6.3$\pm$1.9 & ...\\
$^{13}$CO&$J=13-12$ & 1431.64&  2.18   &10.6$\pm$2.7 &3.5$\pm$1.1 & ...\\
CS & $12-11$ & 587.6 &  9.27     &1.6$\pm$0.2 &3.5$\pm$0.4 & ...\\
HCN & $J=8-7$ & 708.71&  8.04      &3.3$\pm$0.6 & 3.2$\pm$0.5& ...\\
HCN &$J=9-8$ & 797.43&  8.50    &2.0$\pm$0.4 & 1.9$\pm$0.4&0.6$\pm$0.1 \\
HCN &$J=10-9$ & 886.84&  9.50    &3.2$\pm$0.3 & 3.4$\pm$0.5& 1.0$\pm$0.1\\
HCO$^+$&$J=6-5$&535.0&  10.8   &1.5$\pm$0.3 &... &... \\
HCO$^+$& $J=7-6$& 623.96&   8.61    &2.5$\pm$0.3 &1.5$\pm$0.2 &... \\
HCO$^+$&$J=8-7$& 713.26&  8.04    &2.4$\pm$0.3 & 2.3$\pm$0.4&... \\
HF &$J=1-0$& 1232.86&   2.08    &-179$\pm$59 &-160$\pm$40& -12$\pm$3\\
H$_2$O&$1_{1,0}-1_{0,1}$ & 557.18& 10.0     &-5.1$\pm$0.7 &-6.2$\pm$0.7 & -1.0$\pm$0.2\\
H$_2$O& $2_{1,1}-2_{0,2}$ & 751.95&  8.16   &8.5$\pm$1.3 & 4.0$\pm$0.6&1.9$\pm$0.2 \\
H$_2$O& $3_{1,2}-3_{0,3}$& 1097.38&  2.29    &25.7$\pm$8.1 &5.2$\pm$1.6 &... \\
H$_2$O&$1_{1,1}-0_{0,0}$&1113.68 &  2.25    &-163$\pm$55 &-161$\pm$53 & -18$\pm$6\\
H$_2$O&$3_{2,1}-3_{1,2}$& 1163.11 &   2.15    &40$\pm$10 &13.6$\pm$3.9& 2.1$\pm$0.6\\
H$_2$O&$4_{2,2}-4_{1,3}$& 1207.42 &    2.10    &30$\pm$8 &13.2$\pm$3.9 &... \\
H$_2$O&$2_{2,0}-2_{1,1}$& 1228.80 &  2.09    &16.5$\pm$4.4&7.3$\pm$2.3& 1.0$\pm$0.3\\
H$_2$O&$5_{2,3}-5_{1,4}$& 1410.76 &    2.19   &18.7$\pm$5.1 &5.4$\pm$1.4 &... \\
H$_2$O$^+$&$1_{1,1}-0_{0,0}$&  1139.56&   2.19     &-118$\pm$38 &-116$\pm$34 & -7$\pm$2\\
H$_2$S& $2_{2,1}-2_{1,2}$& 505.56&  1.19    &1.5$\pm$0.2 &...&...\\
H$_2$S& $2_{1,2}-1_{0,1}$& 736.35&  8.09    &-9.9$\pm$1.6 &-14.2$\pm$1.7 & -1.2$\pm$0.2\\
{[N\,{\sc ii}]}& & 1462.26& 2.21     &76$\pm$21 &61$\pm$20& 5.0$\pm$1\\
NH &$1_{0}-0_{1}$ & 946.56& 10.2     &-30.0$\pm$7.6 & -32$\pm$9& -4.8$\pm$1.5\\
NH &$1_{2}-0_{1}$ &974.60&  2.80    &-58.8$\pm$15.4 & -57.6$\pm$18.0&-7.4$\pm$2.3\\
NH&$1_{2}-0_{1}$& 974.66& 2.80   &-81.7$\pm$24.1 & -87.9$\pm$25.6& -5.7$\pm$1.6\\
NH&$1_{1}-0_{1}$& 1000.15 &  2.66    &-87.8$\pm$28.0 &-93.4$\pm$26.9& -7.7$\pm$2.2\\
NH$_2$& $2_{1,1}-2_{0,2}$& 648.68&  8.31     &1.3$\pm$0.2 &...&... \\
NH$_2$& $2_{0,2}-1_{1,1}$& 902.42& 9.70  &-9.6$\pm$1.3 &-16.4$\pm$2.3 &-1.9$\pm$0.4 \\
NH$_2$&$1_{1,1}-0_{0,0}$ & 952.79&   2.93 &-52.0$\pm$8.0&-51.1$\pm$6.6& -8.9$\pm$1.8\\
NH$_2$& & 959.75&  2.88  &-43.0$\pm$7.1&-41.7$\pm$5.8 & -8.2$\pm$0.9\\
NH$_2$&$1_{1,1}-0_{0,0}$& 959.68 & 2.88 &-73.0$\pm$8.6&-76.4$\pm$15.0 & -5.5$\pm$1.1\\
NH$_2$&$2_{2}-2_{1}$ & 1383.44&  2.13 &-68.3$\pm$17.8&-56.5$\pm$17.8 & -114$\pm$36\\
NH$_2$&$2_{1}-1_{0}$ & 1444.10&  2.19 &-131$\pm$34&-124$\pm$34&-6.8$\pm$1.9 \\
NH$_2$&$2_{1}-1_{0}$& 1447.72&  2.19   & -115$\pm$36&-119$\pm$30 & ...\\
NH$_3$&$2_{1,-1}-1_{1,1}$& 1215.40 & 2.09 &-77.6$\pm$22.2&-97.8$\pm$30.5& -4.5$\pm$1.4\\
OH$^+$&$1_{0,1}-0_{1,2}$ &909.45 &  9.77&-32.1$\pm$4.3&-29.9$\pm$4.2&-5.4$\pm$0.8 \\
OH$^+$&$1_{2,2}-0_{1,1}$ &972.0&2.81&-104$\pm$33 &-101$\pm$33& -8.0$\pm$2.6\\
OH$^+$&$1_{1,2}-0_{1,2}$&1033.12&  2.51& -129$\pm$33&-121.7$\pm$40.5& -10.0$\pm$3.3\\
\hline
\end{tabular}
\begin{list}{}{}
\item[$^{\mathrm{\star}}$] CO rotational lines detected with PACS.
\end{list}
\end{table*}

\begin{figure*}[!ht]
\centering
\includegraphics[width=12.5cm]{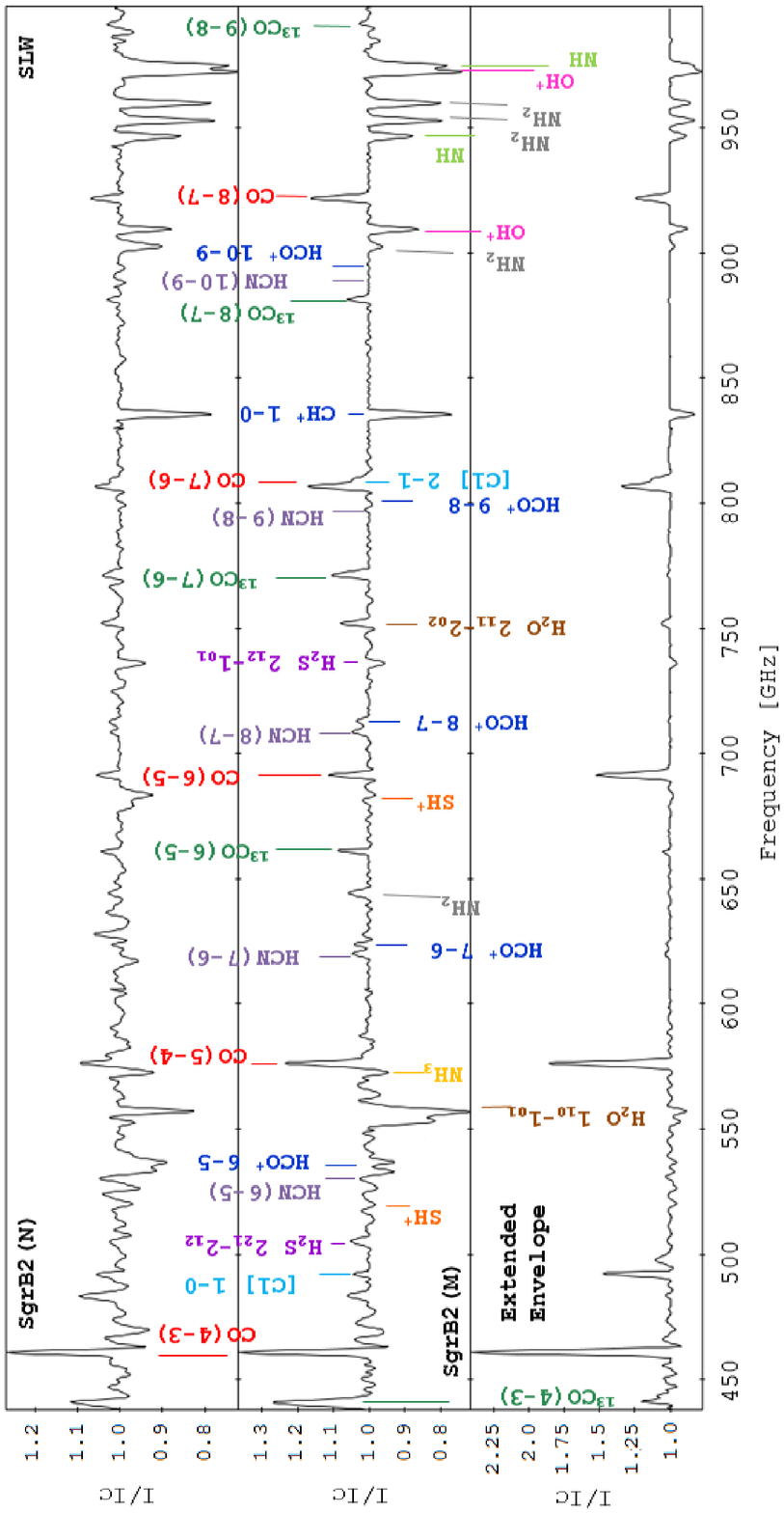}
\caption{Continuum-divided spectra obtained with the SLW detector array, centered at Sgr B2(M), Sgr B2(N), and the extended envelope at $(-1.24',0')$.}
\label{fig10}
\end{figure*}
\begin{figure*}[!ht]
\centering
\includegraphics[width=12.5cm]{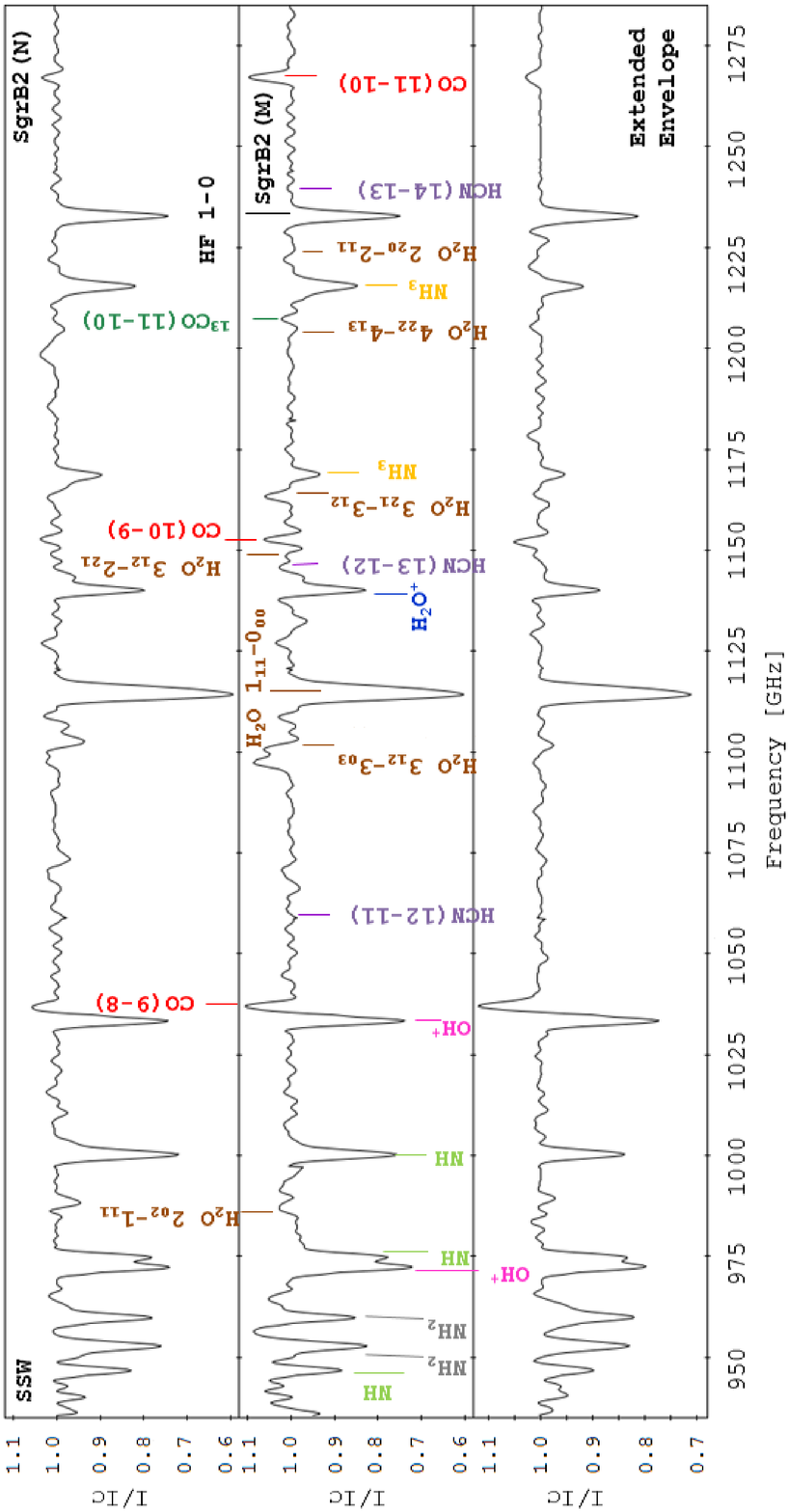}
\caption{Continuum-divided spectra obtained with the SSW detector array, centered in Sgr B2(M), Sgr B2(N), and the extended envelope at $(-1.24',0')$.}
\label{fig11}
\end{figure*}

\par{This provides column densities as high as  $N{\rm (H)}\sim 1.8\times 10^{25}$ cm$^{-2}$, $N{\rm (H)}\sim 9.8\times 10^{24}$ cm$^{-2}$ and $N{\rm (H)}\sim 6.6\times 10^{24}$ cm$^{-2}$ towards Sgr B2(N), Sgr B2(M), and Sgr B2(S), respectively. The region at 0.4 arcmin to the southwest of Sgr B2(M) presents a column density value of  $N{\rm (H)}\sim 1.6\times 10^{25}$ cm$^{-2}$. The dust mass and the total far-infrared luminosity of each core are calculated in Section 3.2.4.
}

\par{The Sgr B2(N) star-forming core with the highest column density may be embedded deeper in the cloud than Sgr B2(M) and Sgr B2(S) \citep{Lang10,Goldsmith87}. The high extinction towards Sgr B2(N) likely explains the lack of VSG emission detected along this core at 24 $\mu$m. In addition, the Sgr B2(M) star-forming core could be more evolved than Sgr B2(N) and located in a more external region on the molecular cloud. This could explain why Sgr B2(M) {presents a higher dust temperature} and a lower hydrogen column density than Sgr B2(N).
}


\subsubsection{Spectral Index}\label{spec-index}


\indent\par{The spectral index ($\beta$) image (Figure~\ref{fig7}c) shows values in the range $\beta \sim 1.1 - 2.4$. Most of the region presents a uniform distribution of $\beta$ with values in the range of 1.7-2.0. The $\beta$ value is slightly lower at the position of the main cores ($\beta \sim 1.5$), with the highest values, $\beta\sim 2.3$, at the southwest of Sgr B2(M) and the lowest values at the north of the cloud ($\beta \sim 1.1$). A spectral index value $\beta \sim 2.0$ is often assumed for crystalline silicates and graphite dust grains in molecular clouds \citep{Hirashita07,Draine84}, while amorphous carbon and aggregates of silicates and graphites in a porous structure are thought to have a spectral index of $\beta \sim 1.0$ \citep{Koike80,Mathis89}. Previous studies of the spectral index of dust emissivity have revealed an anticorrelation between $\beta$ and the dust temperature. \citet{Dupac03} and \citet{Desert08} found evidence of this anticorrelation with measurements from PRONAOS and ARCHEOPS in several different environments through the interstellar medium. More recent studies of T and $\beta$ using {\it Herschel} PACS and SPIRE data \citep{Paradis10,Anderson10,Etxaluze11} confirmed the existence of this anticorrelation at different regions in the Galactic plane. 
}

\par{For Sgr B2 we obtained a flat correlation between T and $\beta$ and did not observe a general anticorrelation (see Figure~\ref{fig9}). In fact, the lowest $\beta$ values are coincident with the lowest temperature values through Sgr B2(N) in the region placed at 1.2 arcmin at the north of Sgr B2(N). This region is chemically rich in HNCO and HOCO$^+$ and may be associated with shocks from cloud-to-cloud collisions \citep{Minh98}.
}

\par{It remains uncertain whether the anticorrelation between $\beta$  and T derived in other regions is real or just a general trend observed either due to the averaging of different values of temperatures along the line of sight or to a large variation of the chemical composition of the grains from region to region \citep{Hirashita07,Butler09}. The anticorrelation can also be due to the fact that an $\chi ^2$ fit produces an artificial anticorrelation because of the degeneracy between T and $\beta$, underestimating $\beta$ and hence overestimating $T_{\rm d}$ \citep{Blain03,Sajina06}. 
}

\subsection{Gas Emission and Absorption}


\indent\par{The two pointed SPIRE FTS observations cover the entire submm band at the position of the main cores (Sgr B2(M) and Sgr B2(N)), at 16 different positions with the SLW, and at 35 positions with the SSW over the extended envelope of the Sgr B2 molecular cloud, providing sparse-sampling spatial distribution of the molecular and the ionized gas through the region. The SPIRE FTS spectral data also allow us to further constrain the physical and the chemical conditions of the gas by comparing the observed spectra with non-LTE radiative models.
}
\par{Figures~\ref{fig10} \&~\ref{fig11} show two continuum-divided spectra centered at the position of Sgr B2(N) and Sgr B2(M), each of them observed with the SLWC3 and SSWD4 detectors. Hence, for optically thin lines, the absorption depths are proportional to the molecular column densities. A third spectrum observed at the extended envelope at a distance of 1.24 arcmin ($\sim$3.0 pc) from Sgr B2(M) to the west is also presented. In order to identify spectral lines accurately, the three spectra were apodized using a Norton-Beer function. This process reduces the $sinc$ sidelobes of the instrumental line shape, allowing us to discriminate real lines from the $sinc$ sidelobes.
}

\par{The entire region shows very strong absorption lines against the background submm continuum emission of rotational lines from the lowest energy levels of a variety of hydride molecules: OH$^+$, CH$^+$, H$_2$O$^+$, H$_2$O, HF, SH$^+$, NH, NH$_2$, and NH$_3$. This is due to the fact that the line of sight towards Sgr B2 passes through the diffuse clouds located in the spiral arms and through the extended envelope of Sgr B2 \citep{Polehampton07,Goicoechea04}. Nevertheless, the spectral resolution is not high enough to discern the exact contribution of each line-of-sight component. Rotationally excited lines, however, arise in the Sgr B2 cloud.
}

\par{The three spectra are also dominated by strong CO rotational lines in emission, especially at the position of Sgr B2(M). Eight rotational CO lines are detected in the three spectra within the frequency range from 445 GHz to 1,300 GHz (from $J=4-3$ to $J=11-10$). 
}
\begin{figure*}[!t]
\includegraphics[width=17.5cm]{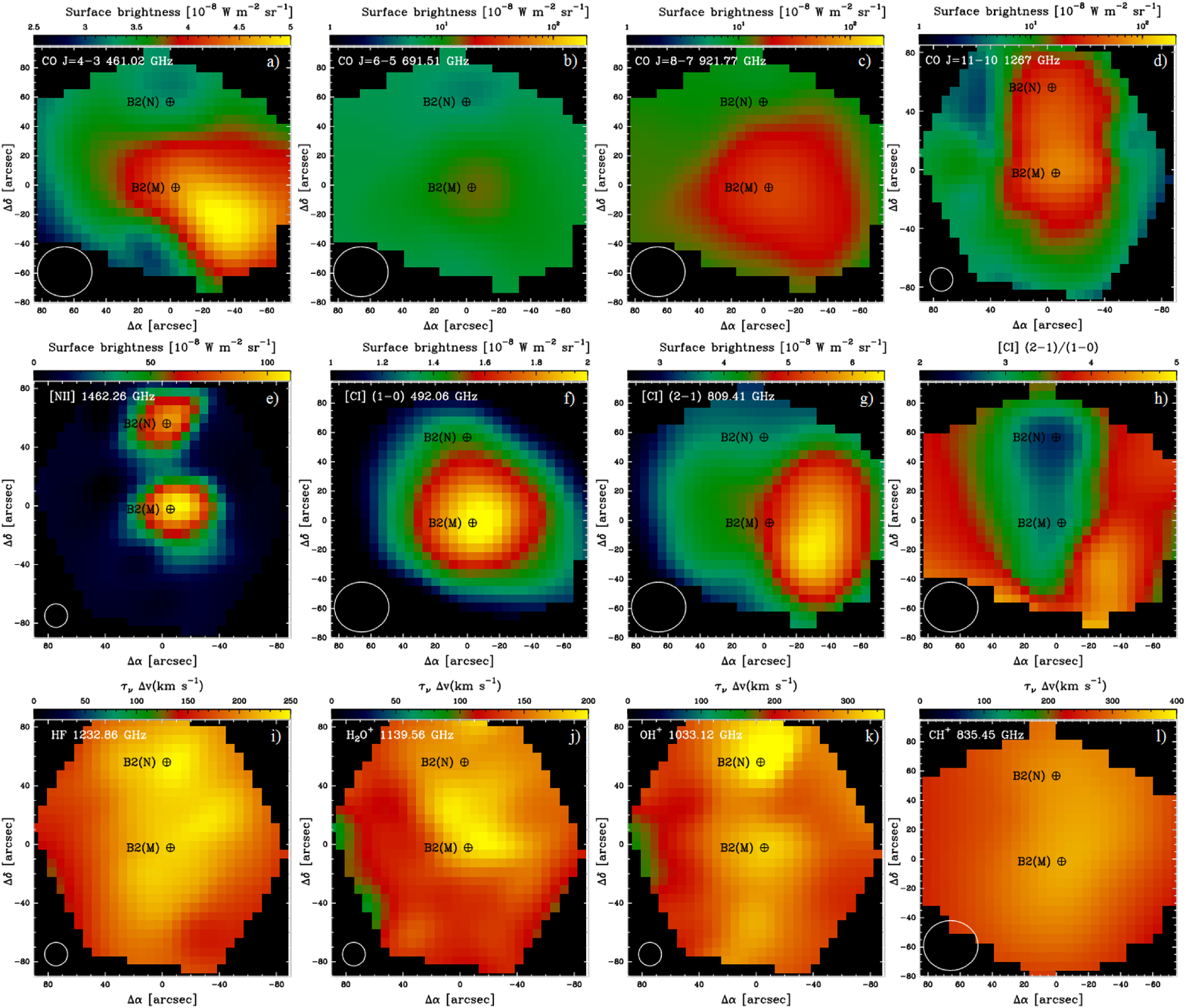}
\centering
\caption{{\it Top:} Maps of the CO $J=4-3$, $J=6-5$, $J=8-7$, and $J=11-10$ lines. {\it Center:} Maps of the emission lines [NII], [CI] (1-0), [CI] (2-1), and [CI] (2-1)/(1-0). {\it Bottom:} Velocity-integrated optical depth maps of the absorption spectral lines: HF, H$_2$O$^+$, OH$^+$, and CH$^+$. The total area of each map is $\sim 3.5\times 3.5$ arcmin$^2$. The maps have a pixel size of 9.5". The maps observed with the SLW detector ($\nu<$ 970 GHz) have a FWHM$\sim$30" and those observed with the SSW detector ($\nu>$ 970 GHz) have a FWHM$\sim$18". All the maps are centered at the position of Sgr B2(M).}
\label{fig12}
\end{figure*}

\par{The spectra also show emission lines from high-density tracers in emission: $^{13}$CO (from $J=4-3$ to $J=9-8$), HCN (from $J=6-5$ to $J=14-13$), and HCO$^+$ (from $J=6-5$ to $J=10-9$). These molecules have larger dipole moments than those of CO. Consequently their rotational lines have higher critical densities ($n{_{\rm cr}\rm (H_2)}\sim 10^6-10^7$ cm$^{-3}$) than those of the CO rotational lines and they therefore probe high-density gas. However, these lines are fainter since their abundance is often $\sim$10$^3$ times less than CO.  Other observed molecular lines are o-H$_2$O (at 556.94 GHz, 1097.36 GHz, 1153.13 GHz, and 1162.91 GHz), and p-H$_2$O (at 752.03 GHz, 987.92 GHz, 1113.34 GHz, and 1207.64 GHz) (Table~\ref{tab1}). 
}
\par{The H$_2$S $2_{1,2}-1_{1,0}$ (736.03 GHz) ground state line is detected in absorption in the three spectra. In contrast, H$_2$S $2_{2,1}-2_{1,2}$ at 505.56 GHz turns up into emission in the Sgr B2(M) spectrum but is undetected in the extended envelope and at Sgr B2(N). H$_2$S is suggested to be an effective shock tracer, probably formed by the passage of C-type shocks and being quickly transformed to SO and  SO$_ 2$ \citep{Pineau93,Charnley97}. \citet{Minh04} also detected the H$_2$S $2_{2,0}-2_{1,1}$ line in emission at 216.7 GHz towards Sgr B2(M) and observed that the emission was negligible towards Sgr B2(N) and other positions in the extended envelope. They suggested that the abundance of H$_2$S is enhanced towards Sgr B2(M) due to the shock processes produced by the star-formation activity taking place in the region, while Sgr B2(N), which is less evolved than Sgr B2(M), has not been sufficiently affected by shocks to produce H$_2$S.  
}
\par{The [C\,{\sc i}] ($^3\rm P_1$ - $^3\rm P_0$) fine structure line at 492.2 GHz is detected in emission. This line is brighter in Sgr B2(M) and Sgr B2(N) than in the extended envelope. The [C\,{\sc i}] ($^3\rm P_2$ - $^3\rm P_1$) line at 809.31 GHz and the CO $J=7-6$ transition at 806.65 GHz are blended in the apodized spectra. Other molecular lines blended are ortho-H$_2$O 620.7 GHz which is blended with HCN $J=7-6$ at 620.3 GHz, and the HCN $J=11-10$ transition at 974.48 GHz blended with the NH absorption line at 974.5 GHz.
}
\par{Sgr B2(M) also presents a clear emission line of NH$_2$ at 648.5 GHz. The line is fainter in Sgr B2(N) and not detected again in the extended envelope. However, the three spectra present strong NH$_2$ absorption lines from the ground level associated with the lower density envelope and with the diffuse clouds along the line of sight. In total, sixty molecular lines in absorption and in emission are resolved in the SPIRE FTS spectra at the position  of Sgr B2(M) and Sgr B2(N).
}
\par{The SSW detector covers a wide range of frequencies between 936 GHz and 1550 GHz. However, we present the spectrum to frequencies up to 1300 GHz because the spectrum is extremely noisy at higher frequencies. Even so, the emission line [N\,{\sc ii}] ($^3\rm P_1$ - $^3\rm P_0$) at 205 $\mu$m is bright enough to be resolved individually, and the analysis and the intensity map of this line are presented in the next section. 
}
\par{Despite the fringing at higher frequencies, the intensity of the emission line CO ($J=12-11$) at 1381.9 GHz was measured, too. The line intensities of all the molecular lines detected with the SPIRE FTS are given in Table~\ref{tab1}.  
}
\par{The detection of the absorption lines from light hydride molecules observed in the SPIRE FTS spectra towards Sgr B2 has been previously reported towards other high-mass star-forming regions using the HIFI instrument onboard {\it Herschel} (H$_2$O$^+$ and OH$^+$: \citet{Ossenkopf10}, \citet{Gerin10}, and \citet{Neufeld10}; NH, NH$_2$, and NH$_3$: \citet{Persson10}; HF: \citet{Neufeld10b}; H$_2$O: \citet{Lis10}; CH$^+$ and SH$^+$: \citet{Falgarone10} and \citet{Godard12}; HCN: \citet{Rolffs10}). 
}

\subsubsection{Mapping Gas Emission and Absorption Lines}\label{mapping-gas}
\indent\par{Figure~\ref{fig12} shows the intensity maps of the line surface brightness of four CO rotational lines: $J=4-3$, $J=6-5$, $J=8-7$, and $J=11-10$, along with the two neutral carbon fine structure transitions, [C\,{\sc i}] ($^3\rm P_1$ - $^3\rm P_0$) at 492 GHz and [C\,{\sc i}] ($^3\rm P_2$ - $^3\rm P_1$) at 809.4 GHz, and the [N\,{\sc ii}] ($^3\rm P_1$ - $^3\rm P_0$) line. We also present the velocity-integrated optical depth maps of several molecules observed in absorption: CH$^+$ $J=1-0$, ortho-H$_2$O$^+$ $1_{1,1}-0_{0,0}$, OH$^+$ $1_{1,2}-0_{1,2}$, and HF $J=1-0$. They were obtained by normalizing the spectra with the continuum and integrating over the velocity ranges. The comparison of these maps allows us to determine the distribution of the different species through the region. 
}
\par
\begin{figure}[!ht]
\resizebox{\hsize}{!}{\includegraphics{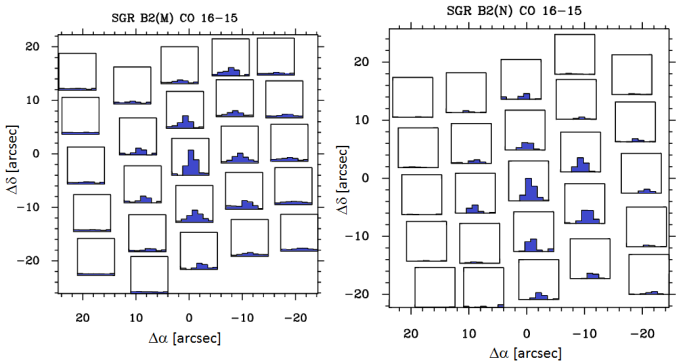}}
\caption{Footprint of the CO $J=16-15$ line observed with PACS towards SgrB2(M) (left) and SgrB2(N) (right).}
\label{fig13}
\end{figure}

{{\it CO:} Figures~\ref{fig12} b) c) and d) show the distribution of three CO rotational lines ($J=6-5$, $J=8-7$, and $J=11-10$) throughout the Sgr B2 molecular cloud. These maps present the same intensity scale. The $J=6-5$ rotational line traces the warm gas, which is smoothly distributed through all the extended envelope with small variations in the intensity scale through the map. Only Sgr B2(M) is faintly resolved at the center of the map. The $J=8-7$ line is again distributed through the entire region, but the line intensity is higher than that of the $J=6-5$ line and the Sgr B2(M) core becomes brighter than in the $J=6-5$ map. The distribution of $J=11-10$ is practically confined to the position of Sgr B2(M) and extends to the north through the location of Sgr B2(N), tracing the distribution of the hot gas towards the main cores. Figure~\ref{fig13} shows the PACS array footprint of the CO $J=16-15$ line observed with PACS, which is centered at the position of Sgr B2(M) and Sgr B2(N). This line peaks clearly at the position of the two main cores.  
}
\par{{\it [C\,{\sc i}] ($^3\rm P_1$ - $^3\rm P_0$):} The first detection of [C\,{\sc i}] at 492 GHz towards Sgr B2(M) was reported by \citet{Sato97}, with observations carried out at the James Clark Maxwell Telescope (JCMT). The [C\,{\sc i}] ($^3\rm P_1$ - $^3\rm P_0$) transition is easily excited since it is only 23 K above ground level and the critical density for collisions with H$_2$ molecules is only $\sim1000$ cm$^{-3}$. Therefore, it is a good tracer of UV-illuminated gas associated with star-formation. Figure~\ref{fig12} f) shows the distribution of [C\,{\sc i}] (1-0) at 492 GHz throughout Sgr B2. The [C\,{\sc i}] (1-0) map shows a peak of emission at the position of Sgr B2(M) decreasing outwards. The integrated intensities of the [C\,{\sc i}] (1-0) are smaller than the CO intensities at every position in the region, indicating that CO is a more important coolant.
}
\par{{\it [C\,{\sc i}] ($^3\rm P_2$ - $^3\rm P_1$) and CO ($J=4-3$):} The [C\,{\sc i}] ($^3\rm P_2$ - $^3\rm P_1$) line was first observed by ~\citet{Jaffe85}. Figures~\ref{fig12} g) and a) show the distributions of [C\,{\sc i}] ($^3\rm P_2$ - $^3\rm P_1$) at 809.4 GHz and CO ($J=4-3$) at 461 GHz, respectively. The distribution of these two lines peaks at the southwest side of Sgr B2(M), differing from the position of the peak of emission of [C\,{\sc i}] (1-0) and the CO rotational lines from transitions $J>5$. In Figure~\ref{fig12} h) the distribution map of the ratio [C\,{\sc i}] (2-1)/(1-0) is presented. The map shows that [C\,{\sc i}] (2-1) intensity is larger than [C\,{\sc i}] (1-0) at every position in the region. The ratio of the two transitions is $\lesssim 3$ towards the cores and presents larger values surrounding the cores, indicating that the emission of [C\,{\sc i}] (2-1) is more extended than that of [C\,{\sc i}] (1-0) within the Sgr B2 region. This suggests that relatively warm UV-illuminated neutral gas surrounds the main star-forming cores.   
}
\par{{\it [N\,{\sc ii}] ($^3\rm P_1$ - $^3\rm P_0$):} Nitrogen has an ionization potential (14.53 eV) larger than that of hydrogen (13.6 eV). Therefore, [N\,{\sc ii}] line arises from ionized gas. Figure~\ref{fig12} e) shows the distribution of [N\,{\sc ii}] mainly confined to the position of Sgr B2(M) and Sgr B2(N), where nitrogen can be ionized by the UV photons from O-type or early B-type stars inside the Sgr B2(M) and Sgr B2(N) H {\sc ii} regions. The surface brightness of the [N\,{\sc ii}] line in Sgr B2(M) is about twice as high as the emission at Sgr B2(N), while the extended envelope surrounding the two main cores presents a negligible emission of [N\,{\sc ii}]. 
}

\par{{\it Absorption lines:} Figures~\ref{fig12} i), j), k), and l) show the velocity-integrated optical depth maps of the spectral lines: HF (1232.5 GHz), H$_2$O$^+$ (1139.5 GHz), OH$^+$ (1033.7 GHz), and CH$^+$ (835.7 GHz). The absorption of these lines is distributed through all the positions uniformly due to the presence of these molecules through the warm extended envelope and the spiral arms along the line of sight. Note that we don't observe significant column density variations at the observed spatial scales. 
}
\subsubsection{CO Emission Rotational Ladder}\label{CO_ladder}
\indent\par{The CO rotational lines provide a fundamental diagnostic of the physical conditions of the molecular gas. As a first approximation, we calculated the CO rotational population diagram with the CO lines (from $J=4-3$ to $J=12-11$) detected by SPIRE FTS (the beam solid angle at the frequency of each CO line is given in Table~\ref{tab1}) at the position of the two main cores, Sgr B2(M) and Sgr B2(N), as well as at 15 different positions through the extended envelope in order to estimate the rotational temperature, $T_{\rm rot}$, which is a good lower limit to $T_{\rm k}$, and the CO beam-averaged column density relative to the SPIRE FTS beams, $N{\rm (CO)}$. We note that three highest CO lines ($J=14-13$, $J=15-14$, and $J=16-15$) detected by PACS. The intensities of the CO lines observed with PACS were integrated in an aperture equivalent to the beam solid angles given in Table~\ref{tab1}, and included in the rotational diagrams of Sgr B2(M) and Sgr B2(N). These high-J lines were not detected by the ISO-LWS telescope, which had a large beam of $\sim 80''$ \citep{Cernicharo06}.
}
\par{The large amount of dust and gas measured towards Sgr B2(M) and Sgr B2(N) implies that the CO rotational lines at far-infrared wavelengths observed with PACS can experience a significant extinction, while this effect may not be so important at the longest submm wavelengths or towards the extended envelope. The way the line intensities are affected by the dust extinction depends on whether the emission is formed at the innermost or at the outermost layers of the cloud. Therefore we compare the rotational temperature values obtained by three different assumptions:}

a) The CO emission is arising from the outermost layers; thus the extinction correction is not considered. 

b) The CO rotational lines are excited in a mixture of gas and dust along the molecular cloud. In this case, the integrated intensities of the CO lines obtained in the spectra of Sgr B2(M) and Sgr B2(N) are corrected for dust as \citep{Pineda10} 

\begin{equation} 
I=I_0\frac{\tau_{\lambda}}{1-{\rm e}^{-\tau_{\lambda}}}
\
;
\
\tau_{\lambda}=\tau_0 \left(\frac{\lambda_0}{\lambda}\right)^{\beta},
\end{equation}
\noindent where $\beta=$2.3 and 2.2 and $\tau_0=$0.94 and 1.35 for Sgr B2(M) and Sgr B2(N), respectively, at $\lambda_0= 250$ $\mu$m. These values were obtained by fitting the spectral continuum (Section 3.2.4.).

c) The CO emission arises from the inner regions of the star-forming cores and a total extinction equivalent to a column density $N{\rm (H_2)}\sim 5\times 10^{24}$ cm$^{-2}$ is assumed. In this case, the extinction correction is given as

\begin{equation} 
I=I_0{\rm e}^{\tau_{\lambda}}
\end{equation}
\begin{figure*}
        \centering
        \begin{subfigure}[]
                \centering
 \includegraphics[width=6.cm]{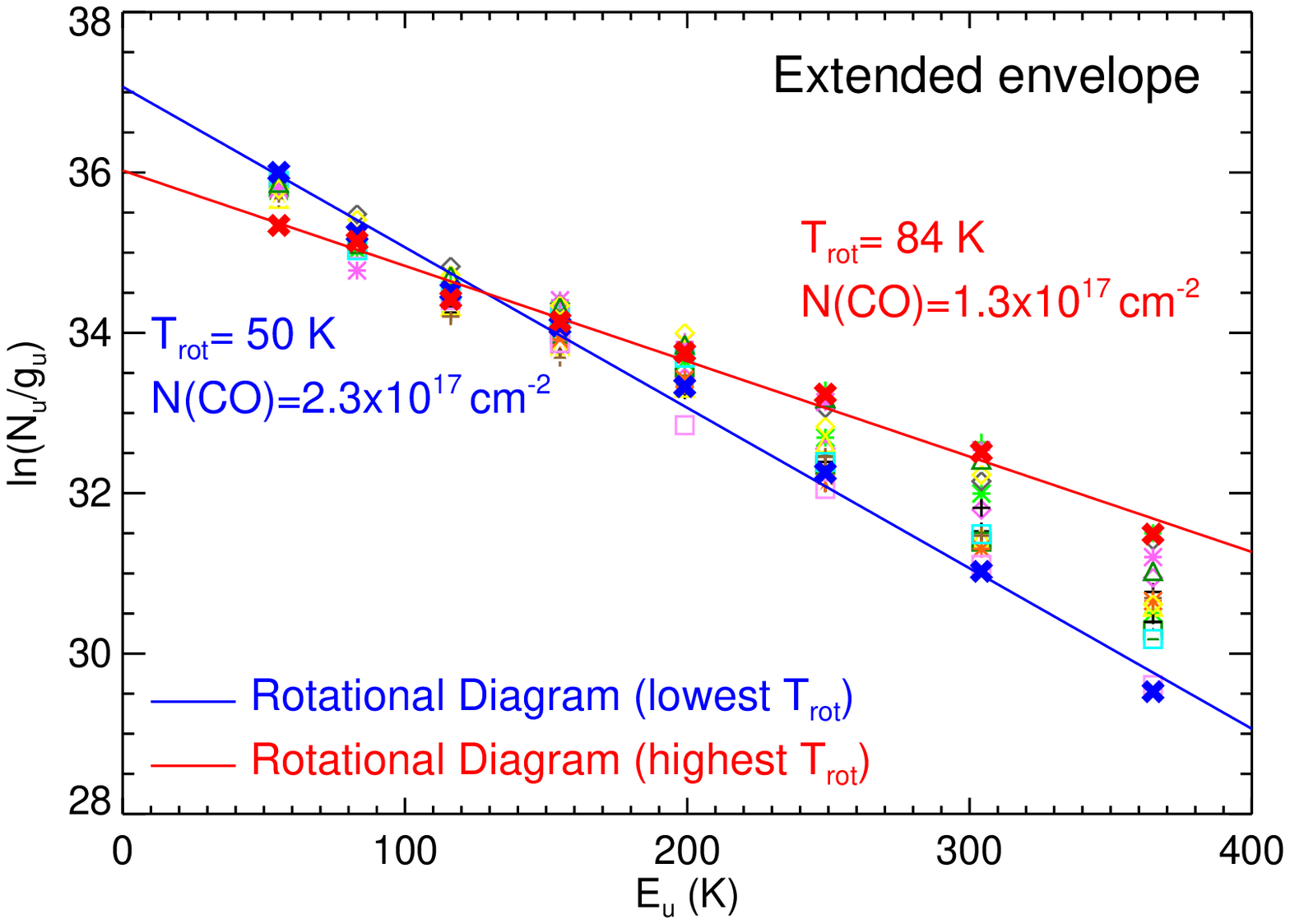}
   \end{subfigure}
        \begin{subfigure}[]
                \centering
  \includegraphics[width=6.cm]{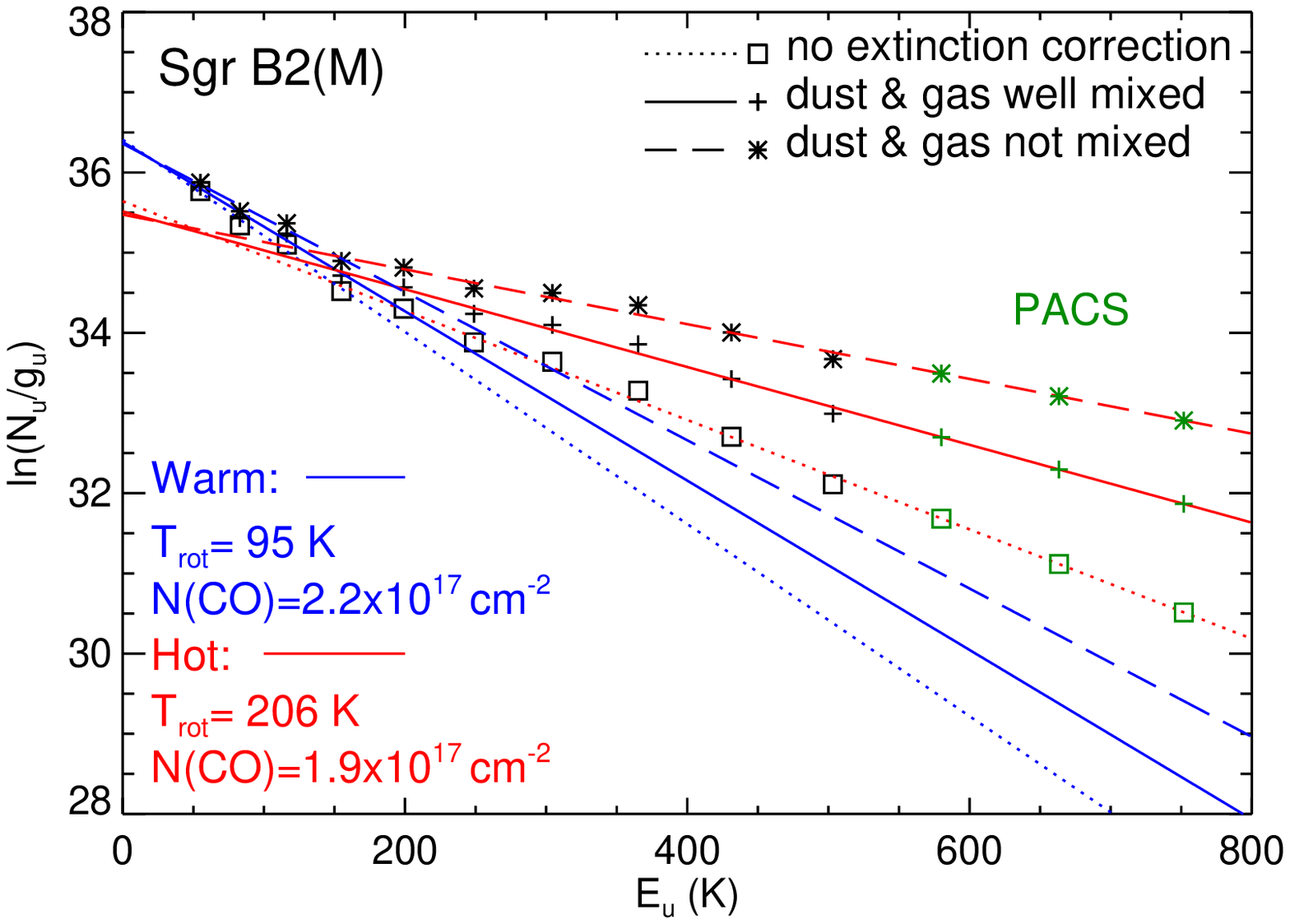}
    \end{subfigure}
        \begin{subfigure}[]
                \centering
   \includegraphics[width=6.cm]{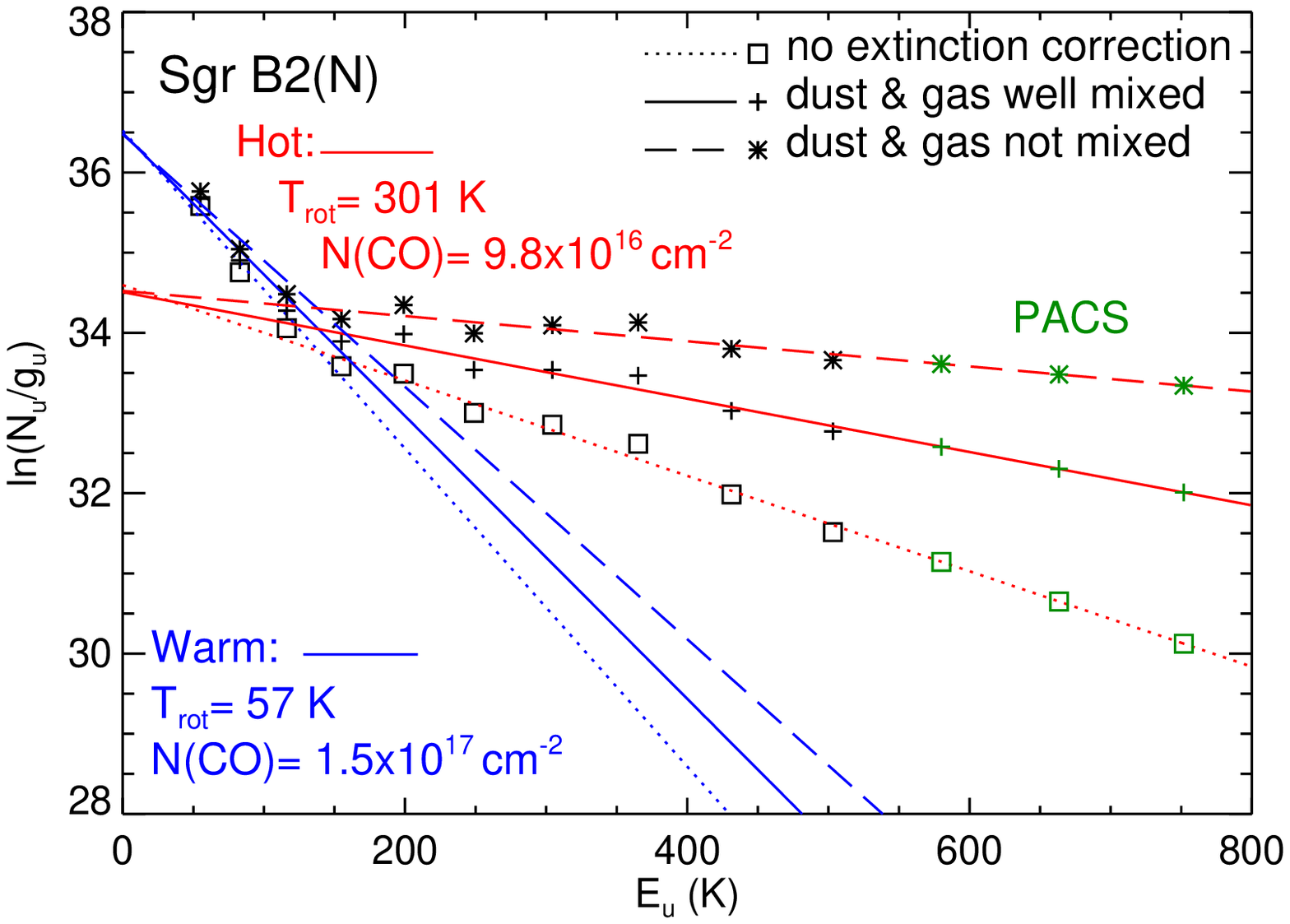}
     \end{subfigure}
        \caption{CO rotational diagrams showing the column density per statistical weight versus the energy of the upper level. (a) The CO rotational diagrams measured at fifteen different positions throughout the warm extended envelope of Sgr B2, limited by the highest $T_{\rm rot}\sim$ 84 K (red line) and the lowest $T_{\rm rot}\sim$ 50 K (blue line). The coordinates and the $T_{\rm rot}$ of each diagram are given in Table~\ref{tab2}. The CO line intensities throughout the envelope are not corrected for dust extinction. The CO rotational diagrams of Sgr B2(M) (b) and Sgr B2(N) (c) are calculated without extinction correction (dashed line), assuming that gas and dust are well mixed (continuum line) and that the CO emission is arising from the innermost region (dotted line). The diagrams suggest the presence of two temperature components, a warm gas component relative to the extended envelope (blue) and a hot gas component associated with the sources themselves (red).}\label{fig14}
\end{figure*}

\begin{table*}
\centering
\caption{$T_{\rm rot}$ (K), $N$(CO) (cm$^{-2}$), and the luminosity ratio $L{\rm (CO)}/L_{\rm FIR}$ throughout the extended envelope and the main cores. Non extinction correction was applied to the CO line intensities in the envelope. Fluxes from Sgr B2(M) and Sgr B2(N) were corrected assuming a mixture of dust and gas.} 
\label{tab2}
\renewcommand{\arraystretch}{1.2}
\begin{tabular}{cccccc}
\hline\hline
 &{$\Delta\alpha$}& {$\Delta\delta$}  & {$T_{\rm rot}$} &{$N$(CO)} &$L{\rm (CO)}/L_{\rm FIR}$\\
&  (arcmin) & (arcmin)& (K) &   ($10^{17}$cm$^{-2}$)&   \\    
\hline
& +0.84& -0.33 & $70\pm4$ & $1.9\pm0.3$& $2.0\times10^{-4}$\\
&+0.89 & -1.11 & $73\pm6$ & $1.4\pm0.1$& $2.3\times10^{-4}$\\
&-0.06 & +1.60& $62\pm4$ & $1.5\pm0.2$& $1.9\times10^{-4}$\\
&-1.61 & +0.58& $52\pm1$ & $1.8\pm0.3$& $1.9\times10^{-4}$\\
&+0.77 & +2.1 & $68\pm3$ & $2.1\pm0.3$& $1.7\times10^{-4}$\\
&+1.53 & +0.94 & $66\pm3$ & $1.4\pm0.1$& $3.5\times10^{-4}$\\
&-0.83 & +1.14 & $64\pm4$ & $1.4\pm0.2$& $2.4\times10^{-4}$\\
&+0.75 & +1.29 & $62\pm3$ & $1.5\pm0.1$& $1.7\times10^{-4}$\\
&+0.80 & +0.45 & $58\pm1$ & $1.7\pm0.3$& $1.6\times10^{-4}$\\
&-0.83 & +0.28 & $63\pm2$ & $2.1\pm0.3$& $1.9\times10^{-4}$\\
&-0.85 & +1.98 & $56\pm1$ & $1.6\pm0.1$& $2.5\times10^{-4}$\\
&-0.65 & -1.30 & $57\pm3$ & $2.0\pm0.2$& $2.2\times10^{-4}$\\
&-0.72 & -0.74 & $77\pm6$ & $2.2\pm0.3$& $1.4\times10^{-4}$\\
&-1.53 &-0.15 & $50\pm1$ & $2.3\pm0.1$& $3.1\times10^{-4}$\\
B2(S)&+0.07 & -0.84 & $84\pm6$ & $1.3\pm0.2$& $1.2\times10^{-4}$\\
B2(M)$_{\rm warm}$&\multirow{2}{*}{+0.00}&\multirow{2}{*}{-0.00} &$95\pm8$ & $2.2\pm0.2$&\multirow{2}{*}{$1.2\times10^{-4}$}\\
B2(M)$_{\rm hot}$&&&$206\pm10$&$1.9\pm0.4$& \\
B2(N)$_{\rm warm}$&\multirow{2}{*}{-0.13} &\multirow{2}{*}{+0.79} &$57\pm6$&$1.4\pm0.2$& \multirow{2}{*}{$0.9\times10^{-4}$}\\
B2(N)$_{\rm hot}$&&&$301\pm7$&$1.0\pm0.1$& \\
\hline
\end{tabular}
\end{table*}

\begin{table*}
\centering
\caption{$T_{\rm rot}$ and $N{\rm (CO)}$ calculated with three different statements of the dust extinction correction at the position of Sgr B2(M) and Sgr B2(N).} 
\label{tab3}
\renewcommand{\arraystretch}{1.2}
\begin{tabular}{@{}lccccccccccc@{}}
\hline\hline
 &\multicolumn{5}{c}{Sgr B2(M)}&&\multicolumn{5}{c}{Sgr B2(N)}   \\
\cmidrule{2-6}\cmidrule{8-12}
& \multicolumn{2}{c}{Warm}&&\multicolumn{2}{c}{Hot}&&\multicolumn{2}{c}{Warm}&&\multicolumn{2}{c}{Hot}\\
\cmidrule{2-3}\cmidrule{5-6}\cmidrule{8-9}\cmidrule{11-12}
& $T_{\rm rot}$& $N{\rm (CO)}$ && $T_{\rm rot}$ & $N{\rm (CO)}$ && $T_{\rm rot}$ & $N{\rm (CO)}$ && $T_{\rm rot}$ & $N{\rm (CO)}$\\ 
& (K) &(10$^{17}$cm$^{-2}$)&& (K) &(10$^{17}$cm$^{-2}$)&& (K) &(10$^{17}$cm$^{-2}$)&& (K) &(10$^{17}$cm$^{-2}$)\\ 
\hline
Not corrected for&\multirow{2}{*}{83$\pm$9}&\multirow{2}{*}{2.0$\pm$0.3}&&\multirow{2}{*}{147$\pm$3}&\multirow{2}{*}{1.6$\pm$0.1}&&\multirow{2}{*}{50$\pm$5}&\multirow{2}{*}{1.4$\pm$0.2}&&\multirow{2}{*}{168$\pm$5}&\multirow{2}{*}{0.6$\pm$0.1}\\ 
dust extinction&&&&&&&&&&& \\
Gas $\&$ dust mixed&{95$\pm$10}&{2.0$\pm$0.4}&&{206$\pm$4}&{2.0$\pm$0.2}&&{57$\pm$9}&{1.5$\pm$0.2}&&{301$\pm$15}&{1.0$\pm$0.2}\\ 
Gas $\&$ dust not mixed&{108$\pm$11}&{2.5$\pm$0.5}&&{293$\pm$9}&{2.5$\pm$0.3}&&{63$\pm$10}&{1.6$\pm$0.3}&&{637$\pm$70}&{1.6$\pm$0.6}\\ 
\hline
\end{tabular}
\end{table*}

\par{The measured extinction in the direction of Sgr B2(M) and Sgr B2(N) is negligible for CO $J\leq 5$, independently of the assumption we take for the location where the CO emission is occurring. However, the extinction calculated assuming that the gas and dust are mixed increases the CO $J=16-15$ rotational line intensity by a factor of 2.7 at Sgr B2(M) and 3.7 at Sgr B2(N). If the CO emission is arising from the inner regions of the star-forming cores, the extinction is much higher raising the intensities by a factor of 10 for Sgr B2(M) and 30 for Sgr B2(N), assuming extended and optically thin emission. 
}
\par{Figure~\ref{fig14} shows the rotational diagrams representing the logarithm of the CO column density per statistical weight, $ln(N_u/g_u)$, for each rotational level versus the energy of the upper level, $E_u$. It is calculated at the 15 different positions over the extended envelope and at the positions of Sgr B2(M) and Sgr B2(N), assuming different extinction correction methods. 
}
\par{The rotational diagrams at the envelope are better fitted with a single-temperature component ({Figure~\ref{fig14}a). The value of the rotational temperature and the CO column density obtained for each position are shown in Table~\ref{tab2}. The errors on $T_{\rm rot}$ and $N{\rm (CO)}$ do not include uncertainties on the calibration or changes in the beam size for each frequency. The errors are just the uncertainty on the fit. The rotational temperatures are in the range $50-84$ K, with the highest value measured near Sgr B2(S). The CO column density varies very slightly over the extended envelope with values in the range $\left (1.3-2.3\right )\times 10^{17}$ cm$^{-2}$. The highest CO column density measured in the extended envelope peaks to the west and southwest of Sgr B2(M), corresponding to the maximum value of emission on the CO ($J=4-3$) and [C\,{\sc i}] ($^3\rm P_2$ - $^3\rm P_1$) maps (Figures~\ref{fig12} a) and g)).
}
\par{We fitted the CO rotational diagrams of Sgr B2(M) and Sgr B2(N) with two temperature components. The $T_{\rm rot}$ and the $N{\rm (CO)}$ measured for each component assuming different dust extinction corrections are given in Table~\ref{tab3}. Sgr B2(N) presents a warm component traced by the low-$J$s ({Figure~\ref{fig14}c) and associated with the extended envelope that encloses the cores. The extinction is small at the lowest $J$, so the rotational temperature values of the warm gas component are essentially unaffected by the assumption we take about the dust extinction.  The temperature values vary from $T_{\rm rot}$= 50 K, assuming the emission is coming from the outer layers, to $T_{\rm rot}$=  63 K if the emission is coming from the inner layers. Alternatively, we could assume that the gas and dust are well mixed, giving $T_{\rm rot}\sim$ 57 K. However, Sgr B2(M) ({Figure~\ref{fig14}b) shows a warm component with a higher rotational temperature, which ranges from $T_{\rm rot}\sim 83$ K if the dust extinction is not corrected to $T_{\rm rot}\sim 108$ K if the CO excitation is occurring in the inner layers. Assuming a mixture of gas and dust, the rotational temperature is $T_{\rm rot}\sim 95$ K. 
}
\par{The rotational diagrams towards the cores present a flatter component that we interpret as a second hot temperature component associated with the star-forming cores themselves. Since this hot component is traced by the mid- and high-$J$, its rotational temperature is very sensitive to the extinction correction method assumed. Sgr B2(M) presents a high temperature component with $T_{\rm rot}=147$ K if the dust extinction is not corrected, $T_{\rm rot}=206$ K assuming that gas and dust are mixed, and $T_{\rm rot}=293$ K if the extinction is calculated over the entire H$_2$ column density. Sgr B2(N) presents a second component with a slightly higher rotational temperature than Sgr B2(M) if the emission was arising from the outer layers $T_{\rm rot}=168$ K. The rotational temperature is higher for a mixture of gas and dust $T_{\rm rot}=301$ K and much higher than that of Sgr B2(M), when the CO emission is considered to arise from the innermost regions, $T_{\rm rot}=700$ K. Previous studies suggest that the kinetic temperature in the Sgr B2(M) and Sgr B2(N) star-forming cores is about a few hundreds K. This is better reproduced when the CO rotational lines are corrected of dust extinction, assuming that dust and gas are well mixed.
}
\begin{figure*}[!t]
        \centering
\begin{subfigure}[]
                \centering
 \includegraphics[width=6cm]{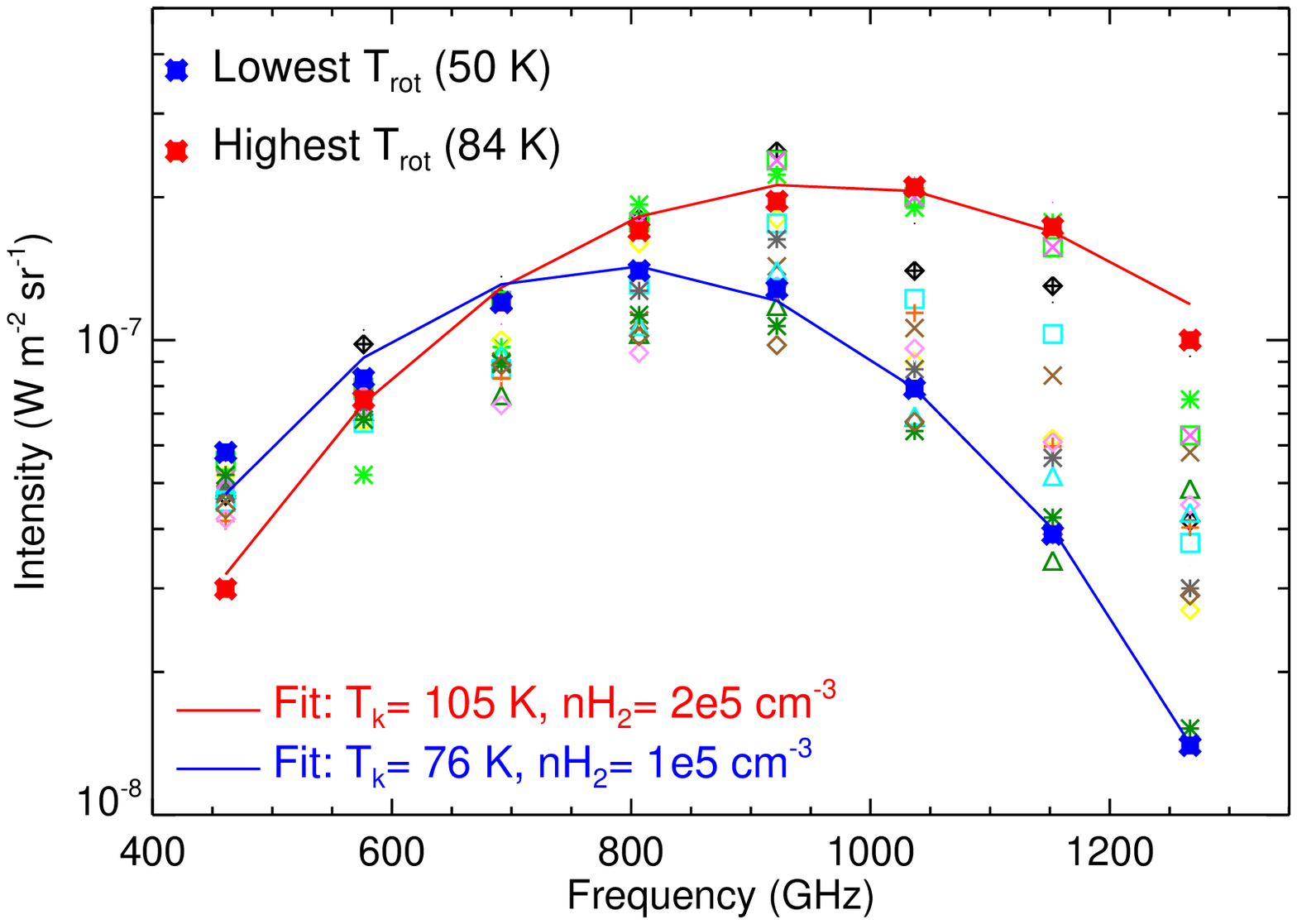}
        \end{subfigure}
        \begin{subfigure}[]
                \centering
 \includegraphics[width=6cm]{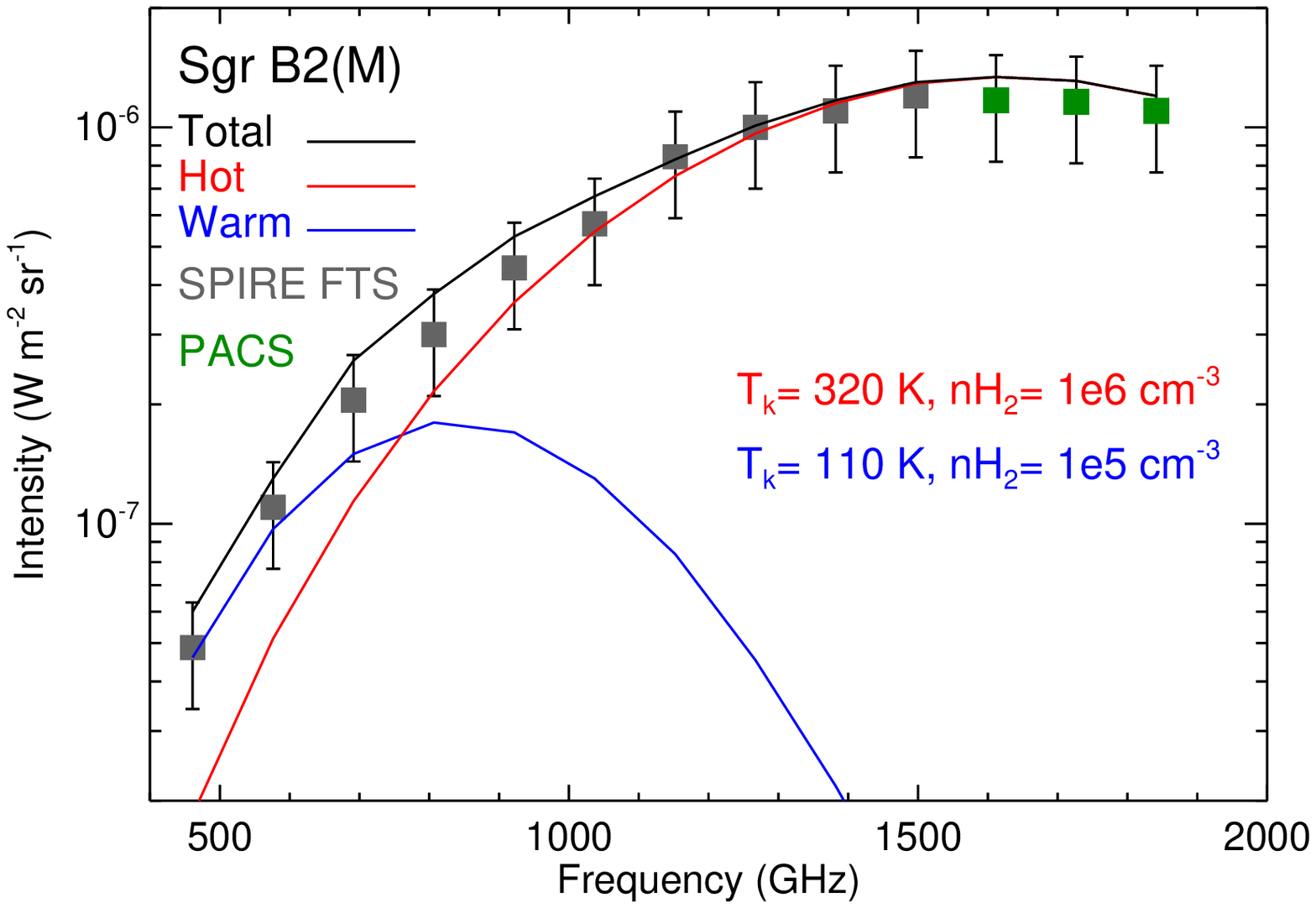}
        \end{subfigure}
        \begin{subfigure}[]
                \centering
  \includegraphics[width=6cm]{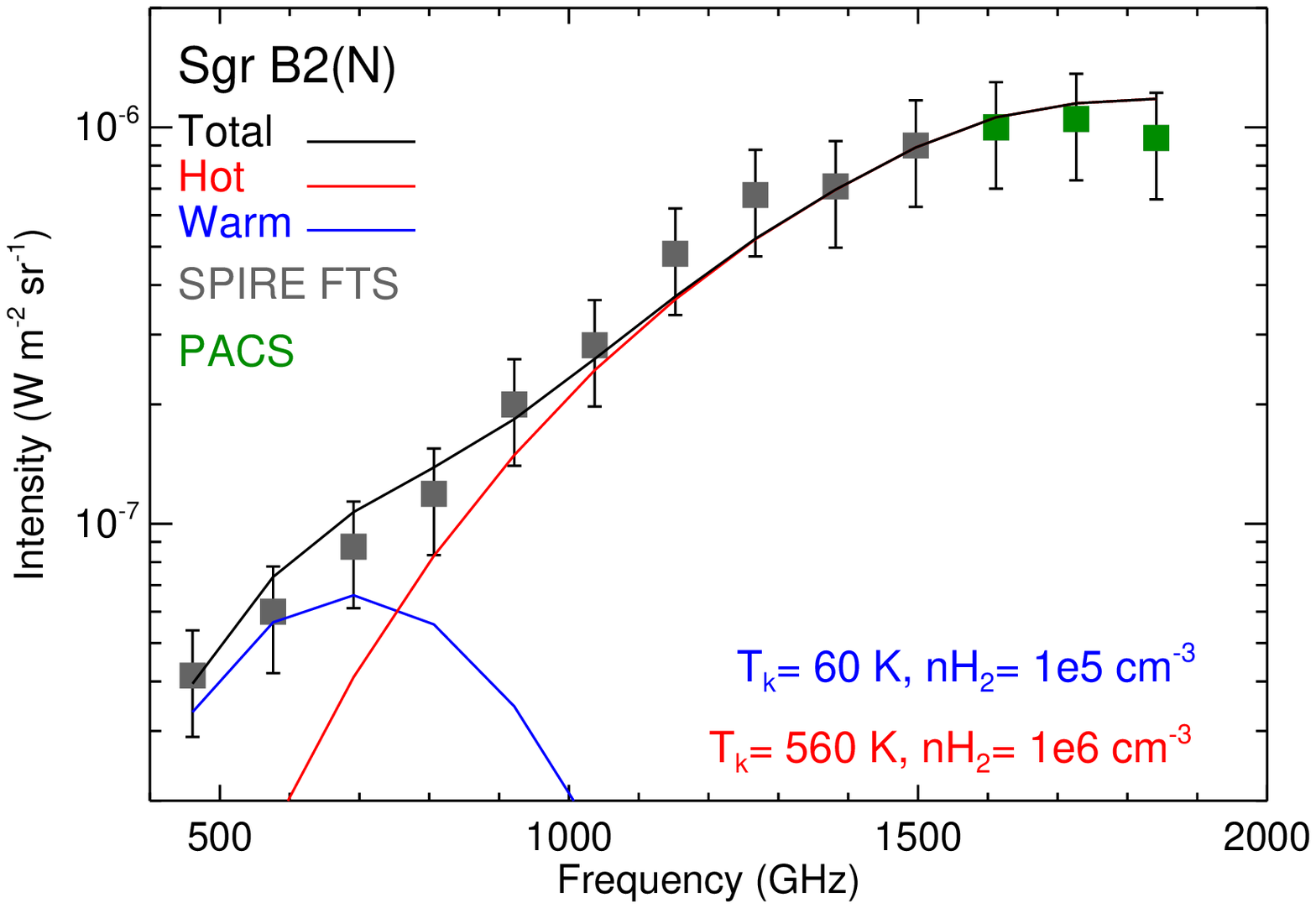}
        \end{subfigure}
                \caption{Integrated CO intensities fitted with non-LTE models. (a) CO emission rotational ladders at fifteen positions throughout the extended envelope. It shows the fits of the CO rotational ladders at the positions of the highest and the lowest $T_{\rm rot}$. (b) and (c) show the CO rotational ladders on Sgr B2(M) (b) and Sgr B2(N) (c) fitted by the non-LTE model obtained (black line). Blue line represents the contribution from the warm gas, which is distributed throughout the extended envelope. Red line is the contribution of the hot gas associated with the main cores. Grey squares represent the fluxes observed with the SPIRE FTS and the green squares are the fluxes observed with the PACS.}\label{fig15}
\end{figure*}
\par{Independently of the extinction correction method we use, the warm components on Sgr B2(M) and Sgr B2(N) present CO beam-averaged column densities of $N{\rm (CO)}\sim 2\times 10^{17}$ and $N{\rm (CO)}\sim 1.5\times 10^{17}$ cm$^{-2}$, respectively. However, the extinction correction statement undertaken affects notably to the $N{\rm (CO)}$ values of the hot component. So, if the dust extinction is not taken into account, the values of total CO column density measured in the direction of the cores are $N{\rm(CO)}= 3.6\times10^{17}$ cm$^{-2}$ for Sgr B2(M), and $N{\rm(CO)}= 2\times10^{17}$ cm$^{-2}$ for Sgr B2(N). In this case, the hot component contribution to the total CO column density is $\sim44 \%$ for Sgr B2(M) and $\sim33\%$ for Sgr B2(N). If the emission is arising from innermost region of the cores, then the total CO column densities are $N{\rm (CO)}= 5\times 10^{17}$ for Sgr B2(M) and $N{\rm (CO)}=3.2\times 10^{17}$ cm$^{-2}$ for Sgr B2(N). Each gas component contributes $50\%$ to the total $N\rm{(CO)}$ on each core.
}
\subsubsection{Modeling the CO Rotational Ladder}\label
{modeling_co}
\par{The molecular hydrogen density measured by previous authors in the main cores of Sgr B2 is $n{\rm (H_2)}\sim 5\times 10^5$ cm$^{-3}$ \citep{Minh98,Goldsmith87}. This density is below or comparable to the critical densities for collisional excitation of the observed CO lines. Hence, we expect the CO excitation to be in a regime where $T_{\rm rot}\lesssim T_{\rm k}$. In order to model the CO intensities towards the Sgr B2 cores, we treated the non-LTE excitation problem and the line opacity effects with MADEX, a large velocity gradient (LVG) code \citep{Cernicharo12}.
}   
\par{In the following, we assume that the two slopes seen in the rotational diagrams of B2(M) and B2(N) correspond to two different gas component. For the LVG modeling, the CO line intensities were corrected of dust extinction assuming that gas and dust are mixed. As a first approximation, the CO column density, N(CO), for each gas component was assumed to be similar to that determined by the CO rotational diagrams. Thus, we searched for a combination of the kinetic temperature, $T_{\rm k}$, and the molecular hydrogen density, $n{\rm (H_2)}$, that reproduces the entire CO ladder from $J=4-3$ to $J=16-15$ (Figure~\ref{fig15}), assuming a line width of $\Delta{\rm v}= 30$ ${\rm km\,s^{-1}}$ (see the observation of the $J=7-6$ CO line with the Caltech Submillimeter Observatory (CSO) in \citet{Cernicharo06}). 
}
\par{{\it Warm Gas Component:} The lowest rotational transitions of CO in the Sgr B2(M) and Sgr B2(N) spectra are associated with the warm gas through the extended envelope. This warm component is better fit with $N{\rm (CO)}\sim 2.2 \times10^{17}$ cm$^{-2}$ and $T_{\rm k}\sim110$ K for Sgr B2(M), and $N{\rm (CO)}
\sim 10^{17}$ cm$^{-2}$ and $T_{\rm k}\sim60$ K for Sgr B2(N), with a molecular hydrogen density $n{\rm (H_2)}\sim 10^5$ cm$^{-3}$ for Sgr B2(M) and Sgr B2(N). 
}
\par{{\it Hot Gas Component:} The higher rotational transitions of CO arise from higher excitation regions, and we interpret them as a second gas component with higher kinetic temperature. This second component is well reproduced with $N{\rm (CO)}=2 \times10^{17}$ cm$^{-2}$ and $T_{\rm k}\sim 320$ K for Sgr B2(M), and $N{\rm (CO)}= 10^{17}$ cm$^{-2}$ and $T_{\rm k}\sim 560$ K for Sgr B2(N), with a molecular hydrogen density $n{\rm (H_2)}\sim 10^6$ cm$^{-3}$ for Sgr B2(M) and Sgr B2(N).
}
\par{The values of the H$_2$ density estimated for the warm extended component are in good agreement with those derived by previous authors ~\citep{Cernicharo06,Greaves94}. Assuming a typical CO abundance of $\chi_{\rm (CO)}\sim 3.0\times 10^{-5}$ ~\citep{Sonnentrucker07,Lis01}, the molecular hydrogen column density traced by the CO is $N{\rm (H_2)}\sim (0.8-1.4)\times 10^{22}$ cm$^{-2}$, which is much smaller than the N(H$_2$) determined from the dust continuum emission ($\sim(5-7)\times 10^{24}$ cm$^{-2}$, Section 3.2.4). This suggests that in Sgr B2(M) and Sgr B2(N), the warm and hot CO emission arises from a very thin layer $\sim(2-5)\times 10^{-3}$ pc and not from the bulk of material in the core.
}
\subsubsection{Mass and Luminosity}\label{Mass}
\par{In addition to the physical properties of the gas, the SPIRE FTS data provide a good measurement of the dust SED. 
}
\par{The SPIRE FTS continuum was previously corrected for the deviation in the continuum level due to the variation of the beam size. The corrected spectra have a FWHM$\sim 40"$. The continuum of both sources, Sgr B2(M) and Sgr B2(N) (Figure~\ref{fig16}), present almost similar slopes, which can be fit with similar spectral index values. However, both spectra cross at $\sim 790$ GHz; the continuum level of Sgr B2(M) is higher than Sgr B2(N) at highest frequencies and becomes lower for shorter frequencies, indicating that the dust in Sgr B2(M) is warmer than in Sgr B2(N).
}
\par{The dust continuum of Sgr B2(N) is better fit by a modified black body curve corresponding to a dust temperature of $T_{\rm d}\sim$ 30 K and an optical depth $\tau_{250}\sim$ 1.35. The spectrum of Sgr B2(M) peaks at higher frequencies than that of Sgr B2(N) and reveals a higher dust temperature $T_{\rm d}\sim$ 37 K and a lower optical depth $\tau_{250}\sim$ 0.94.
}
\par{We derived the dust mass and the molecular hydrogen column density of each core as \citep{Deharveng09,Hildebrand83}
}
\begin{equation}
M_{\rm d}=\frac{F_{250\mu m} d^2}{\kappa_{250\mu m}{\rm B}_{250\mu m}(T_{\rm d})},
\label{eq:Mass}
\end{equation}
\noindent where $F_{250\mu m}$ is the total intensity at 250 $\mu$m, {\it d} is the distance ($\sim8.5$ kpc), ${\rm B}_{250\mu m}(T_{\rm d})$ is the blackbody emission at 250 $\mu$m for a dust temperature $T_{\rm d}$, and $\kappa_{250\mu m}=5.17$ cm$^2$g$^{-1}$ is the dust opacity \citep{Li01}. 

\par{The H$_2$ column density is given as
}
\begin{equation}
N(H_2)=\frac{\chi_{\rm d} F_{250\mu m}}{2.3m_{\rm H} \kappa_{250\mu m} {\rm B}_{250\mu m}(T_{\rm d})\Omega_{{\rm beam}}},
\label{eq:Dens}
\end{equation}
\noindent where we assumed a gas-to-dust ratio $\chi_{\rm d}=100$, $m_{\rm H}$ is the mass of a hydrogen atom, and $\Omega_{{\rm beam}}$ is the beam solid angle. According to Eq.~\ref{eq:Mass} and Eq.~\ref{eq:Dens}, the column density can be written as

\begin{equation}
N(H_2)=\frac{\chi_{\rm d} M_{\rm d}}{2.3 m_{\rm H} d^2 \Omega_{\rm beam}}.
\end{equation}

\par{The derived H$_2$ column densities are $N(\rm H_2)\sim 5\times 10^{24}$ cm$^{-2}$ for Sgr B2(M) and $N(\rm H_2)\sim 7\times 10^{24}$ cm$^{-2}$ for Sgr B2(N). The total masses of the two cores are similar $M_{\rm d}=2300$ M$_{\odot}$ for Sgr B2(M) and $M_{\rm d}=2500$ M$_{\odot}$ for Sgr B2(N). These masses are lower than the total masses determined by \citet{Qin11} using the Submillimeter Array (SMA): $\sim$3300 M$_{\odot}$ for Sgr B2(N) and $\sim$3500 M$_{\odot}$ for Sgr B2(M). They are also higher than the masses calculated by \citet{Gaume90} using H$_2$ column densities based on 1.3 mm dust emission observed by \citet{Goldsmith87}: $\sim$1050 M$_{\odot}$ for Sgr B2(N) and $\sim$1650 M$_{\odot}$ for Sgr B2(M).
}
\begin{figure}[t]
\resizebox{\hsize}{!}{\includegraphics{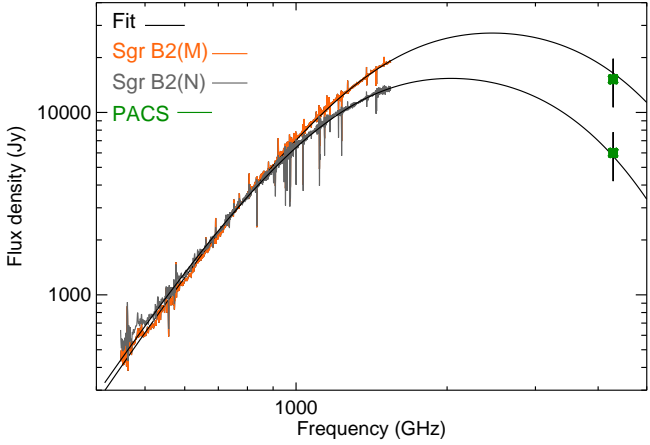}}
\caption{SPIRE FTS spectra of Sgr B2(M) (orange) and Sgr B2(N) (grey). The black line represents the best fit of the continuum with a modified black body curve for each source. Green squares represent the PACS photometric flux at 70 $\mu$m.} 
\label{fig16}
\end{figure}

\par{The total integrated continuum intensity inside the beam FWHM=40" provides luminosities of $L_{\rm FIR}= 5\times10^6$ L$_{\odot}$ for Sgr B2(M), and $L_{\rm FIR}= 1.1\times10^6$ L$_{\odot}$ for Sgr B2(N) (in the range 8 $\mu$m - 1000 $\mu$m), which are in good agreement with those measured by \citet{Goldsmith92}: $7.5\times 10^6$ L$_{\odot}$ for Sgr B2(M) and $10^6$ L$_{\odot}$ for Sgr B2(N). The far-IR luminosity requires several young O-type stars as power sources, which are deeply embedded in the star-forming cores  \citep{Jones08}. The ratio $L_{\rm FIR}/M_{\rm d}$ is indicative of the evolutionary stage of the ongoing high-mass star-formation process, with lower ratios implying earlier stages \citep{Molinari08}. The ratio $L_{\rm FIR}/M_{\rm d}$ is $\sim 2200$ for Sgr B2(M) and $\sim 450$ for Sgr B2(N). This suggests that Sgr B2(N) might be in a more recent stage of star-formation than Sgr B2(M) due to the lower luminosity and the larger amount of dust.  
}
\section{Discussion}\label{discussion}
\indent\par{The Sagittarius B2 star-forming region at $\sim$8.5 kpc is one of the best templates in the Milky Way to understand more distant starburst and ultraluminous infrared galaxies. The spectrum of Sgr B2 observed with the ISO-LWS at the wavelength range 45 to 180 $\mu$m presents similar spectral features to those of Arp 220 \citep{Gonzalez04,Goicoechea04}, showing a very high extinction as well as strong foreground absorption lines of OH, H$_2$O, CH, and [O{\sc i}] towards the nuclei of Arp 220 and the Sgr B2 star-forming cores (B2(M) and B2(N)). The spectrum of M82 starburst Galaxy observed with the ISO-LWS and PACS \citep{Kamenetzky12,Contursi13} presents strong [OI], [OIII], [CII], [NII], and [NIII] fine structure emission lines. However, besides the 119 $\mu$m OH absorption line, no other molecular line was detected in the M82 spectra. At submillimeter wavelengths the SPIRE FTS spectrum of Sgr B2 is again comparable to that of Arp 220 \citep{Rangwala11} and that of M82 \citep{Kamenetzky12}, presenting a bright CO emission ladder from $J= 4-3$ to $J=13-12$ and strong absorption lines from hydrides: OH$^+$, CH$^+$, H$_2$O$^+$, HF, and several nitrogen hydrides. Other molecules detected are HCN and HCO$^+$, along with atomic species such as [C\,{\sc i}] at 492 GHz and 809 GHz and [N\,{\sc ii}] at 205 $\mu$m in emission. However, the far-infrared spectrum of Arp 220 has many features not observed in Sgr B2 and M82, such as strong HCN in absorption and P-Cygni profiles from excited OH$^+$, H$_2$O$^+$, and H$_2$O \citep{Kamenetzky12}. The three sources have a hot gas component in the range of $T_{\rm k}\sim 250-500$ K for M82 and Sgr B2, while the gas is much hotter in Arp 220 ($T_{\rm k}\sim 1350$ K ). Therefore the physical conditions in the Sgr B2 cores are more similar to those observed in M82. The continuum emission of these sources reveals that the properties of the molecular interstellar medium in Arp 220 are quite similar to those of Sgr B2, with a high total hydrogen column density of $\sim 10^{25}$ cm$^{-2}$ and a high dust optical depth $\tau_{100}\sim 5$ \citep{Rangwala11}, while M82 has lower column densities $\sim 10^{22}-10^{23}$ cm$^{-2}$ \citep{Weib01}. 
}
\par{The SED of CO in Sgr B2 is fit by two temperature components: a warm extended gas component ($T_{\rm k}\sim 65$ K) and a hot gas component ($T_{\rm k}\sim 320-560$ K), which  is confined at the position of the two cores. However, if the dust extinction correction is made over the entire H$_2$ column density, the gas temperature obtained is larger than the values determined in previous studies, with $T_{\rm rot}\sim 640$ K at Sgr B2(N). Several mechanisms such as UV photons emitted by OB stars, X-rays (associated with young stellar objects (YSOs), supernova remnants or flares from the Galactic Center), cosmic rays and shocks produced by cloud-cloud collisions, supernova explosions or outflows, can heat the molecular gas in the interstellar medium (ISM), altering the gas physical conditions and modifying its chemistry. Therefore the analysis of line intensities and line ratios of atomic and molecular species provides a good diagnostic to distinguish between the different heating mechanisms. In this work, we calculated the CO-to-FIR luminosity ratio, $L{\rm (CO)}/L_{\rm FIR}$, at 17 different positions within the Sgr B2 molecular cloud (Table~\ref{tab2}). We measured an almost uniform luminosity ratio within the entire region, with values $L{\rm (CO)}/L_{\rm FIR}\sim (1-3)\times 10^{-4}$. This ratio is similar to that predicted by PDR and X-ray dissociation regions (XDR) models but lower than the expected value in shocked gas \citep{Meijerink12}. Throughout the extended envelope ($n{\rm (H_2)}\sim 10^5$ cm$^{-3}$), the observed line ratio ${\rm log[CO(12-11)/CO(5-4)]}$ is in the range [-0.3,-0.9]. This ratio is well reproduced by PDR models for CO from \citet{Wolfire10} (see Fig.~7 at \citet{Rangwala11}), assuming a far-UV radiation field of $G_0\sim$ 10$^3$-10$^4$ \citep{Goicoechea04}. At the position of the Sgr B2 cores, the line ratio is as high as 0.9 for B2(M) and 0.7 for B2(N), and the densities ($n{\rm (H_2)}\sim 10^6$ cm$^{-3}$) determined towards the cores are high enough to reproduce the ratio of the CO lines with the PDR models.
}
\par{The distribution of the highest $J$ CO lines and [N\,{\sc ii}] at 205 $\mu$m are localized at the position of the cores, presenting a strong peak of emission at the position of Sgr B2(M) extending to Sgr B2(N) in the case of the CO. According to the dust emission analysis described in Sections 3.1.3. and 3.2., Sgr B2(M) presents an optical depth at 205 $\mu$m as high as $\tau_{205}\sim 1.5$, with a molecular hydrogen column density as high as $N{\rm (H_2)}\sim 5\times 10^{24}$ cm$^{-2}$ towards the line of sight. The detection of the strong [N\,{\sc ii}] emission line towards a region with such a high dust opacity can only be explained if the cloud surrounding the OB stars located inside the cores is inhomogeneous and clumpy or fragmented enough to allow the percolation of UV photons through the cloud along the line of sight. This clumpy scenario was previously suggested by \citet{Goicoechea04} and \citet{Goldsmith92} for Sgr B2(M) itself and the extended envelope. This is confirmed by our higher angular resolution observations of [N\,{\sc ii}].
}
\par{The leak of UV photons from Sgr B2(M) could explain the higher gas temperature ($T_{\rm k}\sim 110$ K) of the warm gas component at the position of Sgr B2(M) relative to the average $T_{\rm k}\sim 65$ K measured through the extended envelope. Observations of the Sgr B2 cloud using the Submm Array (SMA) \citep{Qin11} show that Sgr B2(M) presents a very fragmented structure with multiple compact cores. In contrast, only two components are resolved in Sgr B2(N), suggesting that Sgr B2(N) is less evolved in agreement with the lower ratio $L_{\rm FIR}/M_{\rm d}$ inferred with the spectral continuum. They also derived molecular hydrogen volume densities of $n{\rm (H_2)}\sim 10^7$ cm$^{-3}$ deeper inside the cores for all submm sources detected in the SMA images. Therefore, in this scenario, it is possible that Sgr B2(N) does not produce enough UV photons or that they do not escape outwards due to a more homogeneous environment, which maintains the warm gas component with a temperature similar to the average value measured along the entire extended envelope, $T_{\rm k}\sim$ 65 K. 
}
\par{The gas density needed to reproduce the mid- and high-$J$ CO rotational lines is $n{\rm (H_2)}\sim 10^6$ cm$^{-3}$ in both star-forming cores, Sgr B2(M) and Sgr B2(N). Nevertheless, the average density of a clumpy cloud is usually less than the density of some of the clumps embedded in the cloud. The clumpiness can explain the detection of high-density tracers such as HCO$^+$ and HCN (previously detected by \citet{Rolffs10} towards Sgr B2(M)), which were observed in emission towards the star-forming cores. CS is also a high-density molecular tracer. \citet{Jones08} observed the emission line CS 2-1 at 97.98 GHz along the line of sight towards Sgr B2. This transition with a critical density $n_{\rm cr}\rm{(H_2)}\sim 5\times 10^5$ cm$^{-3}$ peaks at $\sim$0.4 arcmin to the southwest of Sgr B2(M), coinciding with the position of high optical depth ($\tau_{250} \sim 1.4$) measured on the optical depth map. The existence of clumps with densities higher than $10^6$ cm$^{-3}$ could be responsible for the CO line ratios predicted by the PDR models in the two cores. The gas temperature in Sgr B2(N) is higher than that in Sgr B2(M). \citet{Huttemeister95} also measured higher temperatures in Sgr B2(N) by analyzing seven metastable NH$_3$ inversion transitions and suggested that the high temperatures were consistent with heating by C-shocks. \citet{Ceccarelli02} inferred high gas temperatures $T_{\rm k}\sim 700$ K in the direction of Sgr B2 and suggested that the hot gas created by a shock and associated with Sgr B2 is the cause of the observed absorption lines such as NH$_3$ and H$_2$O.}
\par{The [C\,{\sc i}] ($^3\rm P_2$ - $^3\rm P_1$) and the CO ($J=4-3$) intensity distribution is extended. The peak brightness of the integrated lines of [C\,{\sc i}] (2-1) and CO (J=4-3) transitions occur together at the same position, at $\sim$1 arcmin to the southwest side of Sgr B2(M), implying that these two lines are excited by the same gas component. [C\,{\sc i}] (2-1) is an important coolant in X-ray dissociated regions (XDRs). X-rays can penetrate much deeper into the clouds than the UV photons, so X-ray dissociated gas has different thermal and chemical structures than that irradiated by UV photons \citep{Maloney96}. \citet{Murakami01} detected a diffuse X-ray emission in the southwest side of Sgr B2 with a concave shape pointing towards the Galactic Center; however, no luminous X-ray source is found near the Sgr B2 cloud. \citet{Koyama96} and \citet{Murakami00} suggested that the X-ray sources for producing the diffuse X-ray emission at the southwest of the Sgr B2(M) should be at the Galactic Center coincident with Sgr A$^*$, at a projected distance of $\sim 100$ pc from Sgr B2. The peak of the emission in the maps of the [C\,{\sc i}] (2-1) line and the CO ($J=4-3$) rotational line is also coincident with the position of largest CO column density, $N{\rm (CO)}\sim 2.3\times 10^{17}$ cm$^{-2}$, determined from the CO rotational diagrams along the entire extended envelope (Table~\ref{tab2}). PDR-XDR models \citep{Spaans08} predict that the flux distribution over the CO rotational ladder is different for PDRs and XDRs with high-J ($J>10$) CO line's intensity ratios in XDRs always larger than the corresponding ratios in PDRs \citep{Meijerink06}. The observed values of the CO line ratio CO(10-9)/CO(7-6) along the extended envelope are in the range 0.3 to 1, with the highest values ($>0.8$) distributed through the southwest coinciding with the region where the diffuse X-rays are emitting and with the position of the peak of emission on the maps of the [C\,{\sc i}] (2-1) line and the CO ($J=4-3$) rotational line. 
}
\par{The X-ray luminosity toward Sgr B2 is $L_{\rm X}\sim 10^{35}$ ergs$\ $s$^{-1}$ \citep{Murakami01}, which corresponds to a low X-ray flux incident in molecular gas in Sgr B2 of $F_{\rm X}\sim 10^{-3}$ ergs$\ $cm$^{-2}$$\ $s$^{-1}$ (assuming 1 pc distance from the X-ray source). The X-ray flux is two orders of magnitude lower than the X-ray flux necessary to reproduce the observed CO line ratios according to the XDR models \citep{Meijerink06}. As a result, the contribution of the diffuse X-ray emission to the heating of the gas in the molecular cloud may be negligible. The observed CO(10-9)/CO(7-6) ratio through the southwest of the extended envelope is well reproduced by PDR models at a density $n{\rm (H_2)}= 10^5$ cm$^{-3}$ with $G_0=10^3-10^4$ \citep{Meijerink06}.  
}
\section{Summary and conclusions}\label{summary}
\par{We have presented a detailed analysis of the gas and the dust properties in the Sgr B2 molecular cloud using {\it Herschel} PACS and SPIRE data complemented with Spitzer MIPS data.
}
\par{$-$ SPIRE  submm images trace the dust emission through the main massive star-forming cores and reveal the extended envelope, which extends to the north, west and south. The contribution of the hot VSG population traced by the Spitzer 24 $\mu$m emission is mainly distributed throughout the H\,{\sc ii} regions that exist far from the main cores.  The VSG contribution is very low at the position of the cores, which is likely due to the high dust extinction towards them. The dust temperature image peaks at the position of Sgr B2(M), while the optical depth image peaks at the position of Sgr B2(N).
}
\par{$-$ The SPIRE FTS spectra resolve more than sixty molecular and atomic lines in the submm range $\sim$440-1600 GHz towards the Sgr B2 molecular cloud. The spectra towards the star-forming cores are characterized by strong CO line emission ($J=4-16$), emission lines from high-density tracers (HCN, HCO$^+$ and H$_2$S), emission from ionized gas [N\,{\sc ii}]  205 $\mu$m, and a
large number of absorption lines from hydride molecules (OH$^+$, H$_2$O$^+$, H$_2$O, CH$^+$, CH, SH$^+$, HF, NH, NH$_2$, and NH$_3$).
}
\par{$-$ Maps of the integrated intensity of mid-$J$ CO rotational lines, and of [C\,{\sc i}] and [N\,{\sc ii}] 205$\mu$m fine structure lines are presented together with the maps of HF, H$_2$O$^+$, OH$^+$, and CH$^+$ absorption lines. The CO line $J=6-5$ traces the warm gas, which is smoothly distributed throughout the extended envelope, and only Sgr B2(M) is slightly resolved at the center of the image. As $J$ increases, the main star-forming cores become more prominent and the $J>10$ lines are much stronger towards Sgr B2(M) and Sgr B2(N), where the hot gas is confined, than towards the extended envelope. 
}
\par{$-$ The rotational population diagrams of CO suggest the presence of two different gas temperature components: an extended warm component with $T_{\rm rot}\sim (50-84)$ K associated with the extended envelope and a hotter component with $T_{\rm rot}\sim 200$ K and $T_{\rm rot}\sim 300$ K towards the B2(M) and B2(N) cores respectively. The CO beam-averaged column density presents very weak variations within the whole extended envelope, with values in the range $N{\rm (CO)}\sim \left(1.3-2.3\right)\times 10^{17}$ cm$^{-2}$
}
\par{$-$ Non-LTE models of the CO excitation constrain a gas temperature $T_{\rm k}\sim 320$ K and  $T_{\rm k}\sim 560$ K in Sgr B2(M) and Sgr B2(N) respectively, along with an averaged gas density $n{\rm (H_2)}\sim 10^6$ cm$^{-3}$ in both sources. The gas temperature is substantially higher than the dust temperature determined from the photometric images ($T_{\rm d}\sim  20-30$ K). 
}
\par{$-$ The total integrated continuum intensity indicates luminosities of $L_{\rm FIR}{\rm (SgrB2(M))}= 5\times10^6$ L$_{\odot}$ and $L_{\rm FIR}{\rm (SgrB2(N))}= 1.1\times10^6$ L$_{\odot}$ in an area of $\sim$0.8 pc around each source. The total dust masses are $M_{\rm d}{\rm (SgrB2(M))}=2300$ M$_{\odot}$ and $M_{\rm d}{\rm (SgrB2(N))}= 2500$ M$_{\odot}$. PDR models are able to explain the observed homogeneous luminosity ratio through the cores and the extended envelope ($L{\rm (CO)}/L_{\rm FIR}\sim (1-3)\times 10^{-4}$) and the measured CO intensity line ratios along the extended envelope, indicating that UV photons likely dominate the heating of the molecular gas. A minor contribution from large-scale shocks and cosmic rays cannot be discarded. Shocks from outflows associated with the star-formation activity that takes place in the two star-forming cores, together with a strong UV-radiation ($G_0\sim 10^3-10^4$), likely heats the molecular gas in the cores to temperatures $T_k\gtrsim 300$ K.
}
\begin{acknowledgement}
We are very grateful to the referee, Dr. Paul Goldsmith, for the very valuable comments and suggestions that helped to improve the clarity and the quality of this paper. We thank the HiGAL consortium for releasing the Sgr B2 data. We thank ASTROMADRID for funding support through the grant S2009ESP-1496, the consolider programme ASTROMOL: CSD2009-00038 and the Spanish MINECO (grants AYA2009-07304 and AYA2012-32032). HIPE is a joint development by the {\it Herschel} Science Ground Segment Consortium consisting of ESA, the NASA {\it Herschel} Science Center, and the HIFI, PACS, and SPIRE consortia. SPIRE has been developed by a consortium of institutes led by Cardiff University (UK) and including University of Lethbridge (Canada); NAOC (China); CEA, LAM (France); IFSI, Univ. Padua (Italy); IAC (Spain); Stockholm Observatory (Sweden); Imperial College London, RAL, UCL-MSSL, UKATC, Univ. Sussex (UK); and Caltech, JPL, NHSC, Univ. Colorado (USA). This development has been supported by national funding agencies: CSA (Canada); NAOC (China); CEA, CNES, CNRS (France); ASI (Italy); MCINN (Spain); SNSB (Sweden); STFC (UK); and NASA (USA).

\end{acknowledgement}

\end{document}